\documentclass[12pt,eqsecnum,preprint,flushrt]{aastex}
\usepackage{amsmath}
\usepackage{graphicx}

\newcommand{\be}{\begin{equation}}
\newcommand{\ee}{\end{equation}}
\newcommand{\msun}{{ M_\odot }}  
\newcommand{\yr}{{ \rm yr }} 
\newcommand{\AU}{{ \rm AU }}  
\newcommand{\cm}{{ \rm cm }} 
\newcommand{\s}{{ \rm s }}

\renewcommand\footnotemark{}

\begin{document}

\title{VERTICAL STRUCTURE OF MAGNETIZED ACCRETION DISKS AROUND YOUNG STARS}

\author {S. Lizano$^1$, C. Tapia$^1$, Y. Boehler$^{1,2}$, and P. D'Alessio$^{ \dagger}$}

\affil{$^1$Instituto de Radioastronom{\'i}a y Astrof{\'i}sica, UNAM, Apartado
Postal 3-72, 58089 Morelia, Michoac\'an, M\'exico }
\affil{$^2$Department of Physics and Astronomy
Rice University, 6100 Main Street, Houston, TX, 77005, USA}

\altaffiltext{$^\dagger$}{Untimely deceased, November 14, 2013} 

\begin{abstract} 
We model the vertical structure of magnetized accretion disks subject to viscous and
resistive heating, and irradiation by the central star. We apply our formalism to the radial structure of 
 magnetized accretion disks threaded by a poloidal magnetic field dragged during the process of star
 formation developed by Shu and coworkers.
We consider disks around  low mass protostars, T Tauri, and FU Orionis stars. We consider two
levels of disk magnetization, $\lambda_{sys} = 4$ (strongly magnetized disks), and $\lambda_{sys} = 12$
(weakly magnetized disks). The rotation rates of strongly magnetized disks have large deviations from Keplerian 
rotation. In these models, resistive heating dominates the thermal structure for the FU Ori disk. The T Tauri disk
is very thin and cold because it is strongly compressed by magnetic pressure; it may be too thin
compared with observations.  Instead, in the weakly magnetized disks, rotation velocities are close to
Keplerian, and resistive heating is always less than 7\% of the viscous heating. In these models, the T Tauri disk has a larger
aspect ratio, consistent with that inferred from observations.
All the disks have spatially extended hot atmospheres where the irradiation flux is absorbed, although
most of the mass ($\sim 90-95$ \%)  is in the disk midplane.
With the advent of ALMA one expects direct measurements of magnetic
fields and their morphology at disk scales. It will then be possible to determine the 
mass-to-flux ratio of magnetized accretion disks around young stars, 
an essential parameter for their structure and evolution. Our models contribute to the understanding
of the vertical structure and emission of these disks.

\end{abstract} 


\section{Introduction}

During the last two decades disks around young stars have
been observed from optical to radio wavelengths (e.g., Wiliams \& Cieza 2011) and their
physical properties have  been inferred successfully through models
of their vertical structure and emission (e.g.,  Chiang \& Goldreich 1997; D'Alessio 
et al. 1998). These models consider
the viscous heating of the gas and the heating of the disk surface by 
irradiation from the central star. 
Stellar irradiation is an important source of disk heating during the T Tauri phase, while for 
embedded Class 0 sources the envelope irradiation is also important (e.g., D'Alessio, Calvet \& Hartmann 1997).
Other heating mechanisms have been considered like the accretion
shocks expected during disk formation (Neufeld \& Hollenbach 1994);
bending wave dissipation produced by  a stellar dipole/disk misalignment 
(Lubow \& Pringle 2010); and cosmic rays and X rays produced by the stellar magnetospheres 
(e.g., Igea \& Glassgold 1999; Glassgold, Galli \& Padovani 2013). The effect of
high energy radiation (X rays, UV, and FU) on the evaporation of the disks
around low mass stars has also been studied (e.g., Font et al 2004; Gorti \& Hollenbach 2009;
Gorti, Dullemond \& Hollenbach 2009; Owen, Clarke \& Ercolano 2012).

In accretion disks, turbulence produced by the magnetorotational 
instability (MRI; e.g., Balbus \& Haley 1998) is believed to be the mechanism responsible 
for the anomalous viscosity that allows the inward transport of mass and the outward transport of angular momentum.
Anomalous values are required to explain the fast timescales of 
disk evolution, of the order of  5 - 10 Myr (Strom et al. 1989, Haisch, Lada \& Lada 2001; 
Sicilia-Aguilar et al. 2006, Hern\'andez, Hartmann \& Megeath 2007; Bell et al. 2013).
Because comic rays and X rays can only penetrate mass  column densities
$\Sigma \sim 50 - 100 {\rm g / cm^2}$ (e.g., Umebayashi \& Nakano 1981; 
Igea \& Glassgold 1999; Padovani et al. 2009) 
 the dense gas at the disk midplane is expected to 
be weakly ionized. 
The lack of ionization would produce midplane ``dead zones'' unable to sustain the
MRI, although accretion could still occur in surface layers (e.g., Gammie 1996).

Numerical  shearing box simulations of the MRI with a vertical net flux, as expected from the
poloidal field dragged in during the process of disk formation, have been studied by several authors
(e.g., Suzuki \& Inutsuka 2009, Suzuki, Muto \& Inutzuka 2010, and Fromang et al. 2013;  Bai \& Stone 2013a).
They find that the behavior of the MRI turbulence depends on the ratio of gas to magnetic pressure, such
that the height-integrated mass-weighted Shakura-Sunyaev parameter $\alpha$ is larger than 1
for magnetically dominated disks. In these simulations, 
a disk outflow is launched but the large scale field has no permanent bending direction.
Bai \& Stone (2013b) found that ambipolar diffusion (AD) suppresses 
the MRI: the simulations develop a laminar flow with a strong disk wind that carries
away the angular momentum and drives disk accretion. { Recently, several authors have made simulations
that include the non-ideal MHD effects: Ohmic resistivity, ambipolar diffusion and the Hall effect 
(e.g., Lesur 2014;  Bai 2015; Simon 2015). 
They find that a magnetocentrifugal wind is launched when the vertical field is not too weak and that the 
Hall effect leads to strong Maxwell stresses when the magnetic field is aligned with the disk rotation. 
Lesur et al. (2014) find that large accretion rates can be produced in the aligned case. }
Nevertheless,  global simulations are needed that 
include the effect of the back reaction of the magnetic field on the flow since sub-Keplerian
rotation may hinder the ejection of disk winds (Shu et al 2008). In addition,
Van Ballegooijen (1989) and Lubow et al. (1994) showed that the radial transport of magnetic flux
in magnetized accretion disks depends largely on the ratio of turbulent
viscosity and resistivity.  In particular, the dragging of field lines by accretion is balanced
by the outward diffusion only if the the Prandtl number is
 $P_T = \nu/\eta    \sim 1/ A  > 1 $, where $A$ is the disk aspect ratio.
Guilet \& Ogilvie (2013; 2014) showed that for weak magnetic fields the
advection velocity of the magnetic flux  depends also on the vertical variation 
of the diffusion coefficients, and protoplanetary disks with $P_T \sim 1$ would evolve to a configuration
with a ratio of thermal to magnetic pressure $\beta \sim 10^4 - 10^7$. 
Recent numerical simulations in the shear box approximation have measured the resistivity associated with 
the MRI turbulence and find a  Prandtl number $P_T \sim 1$ (e.g., Fromang \& Stone 2009;
Guan \& Gammie 2009; Lesur \& Longaretti 2009). Nevertheless, by the nature of the
shearing box,  these studies cannot address the magnetic field topology expected in 
magnetized accretion disks. Again, global simulations are needed to study this problem,
including the back reaction of the Lorentz force on the gas. 

The hourglass signature of magnetic fields dragged during the phase of gravitational collapse
has been found with SMA observations of dust polarized emission, for example,
 in the low mass star forming region NGC 1333 IRS5 (Girart, Rao \& Marrone 2006) and
in the high mass star forming region W51 (Tang et al. 2009). 
With the beginning of ALMA operations, one can soon expect the direct detection of Zeeman 
splitting in molecules like CN  and polarized dust emission at the scales of  accretion disks 
{ as has been found recently in a few Young Stellar Objetcs (YSO's) with the SMA and CARMA 
(Rao et al. 2014; Stephens et al. 2014; Segura-Cox et al. 2015). }
Thus, at this time, it is necessary to consider models of accretions disks that 
include the effect of the magnetic field on their structure and emission.
In magnetized disks, both viscous and resistive diffusion are
needed for the gas to lose angular momentum and cross field lines to accrete onto the central star.
Therefore, for magnetized disks, heating by resistive dissipation is a necessary ingredient in the
modeling of the disk structure and evolution.

Analytic models of the radial structure of magnetized disks threaded by a poloidal magnetic field 
dragged into the system by the star formation process have been studied by 
Shu et al. (2007; hereafter S07). They 
showed that because the poloidal field is bent by the accretion flow, 
the magnetic tension produces sub-Keplerian rotation of the gas.  
In their models, resistive dissipation competes with 
viscous heating for disks expected around T Tauri and FU Ori stars.
Also, the disks become more magnetized with time
as the mass accretes on the star and the magnetic field is left behind.
The Toomre stability parameter is modified by two
opposing effects: magnetic pressure and tension support the gas
against gravitational collapse, but sub-Keplerian rotation makes
the gas locally more unstable (Lizano et al. 2010).   In disks around young stars,
the magnetic Toomre stability parameter is larger than its
nonmagnetic counterpart, thus, stable magnetized disks can be more massive that nonmagnetic disks.
Also, the region of instability is pushed out to large radii, making it more
difficult to form giant planets via gravitational instability. In addition,
planet migration is accelerated because the protoplanets move at Keplerian speeds and 
experience a headwind against the slower sub-Keplerian gas (Adams, Cai \& Lizano 2009).

The poloidal field that threads the disk is
dragged from the parent core during the phase of gravitational collapse
and disk formation. Galli et al. (2006) found that magnetic field has to be 
dissipated during the phase of gravitational collapse to prevent the catastrophic 
 braking produced by magnetic torques and allow the 
formation of a rotationally supported disk (RSD;  see also Shu et al. 2006).  
Misalignment between the magnetic and rotation axis, as observed between polarization
vectors in dense cores at scales of 1000 AU and outflows (Hull et al. 2013), can alleviate 
this problem but magnetic field dissipation is still required to form RSDs
(e.g., Hennebelle \& Ciardi 2009).
Several numerical simulations have
been carried out recently to study the conditions for disk formation in magnetized cores
(e.g., Machida, Inutsuka \& Matsumoto  2011, 2014; Li, Kransopolsky \& Shang 2011, 2013; Joos,
Hennebelle \& Ciardi 2012; Santos-Lima, de Gouveia Dal Pino \& Lazarian 2013;   
Seifried et al. 2012, Li et al. 2014; for a review see Lizano \& Galli 2015). 
From initial core values of the 
dimensionless mass-to-flux ratio, $\lambda_{\rm core} = M/2 \pi G \Phi \sim 1 - 4 $,
where $M$ is the mass of the core and $\Phi$ is the magnetic flux,
one expects, after some field dissipation, higher mass-to-flux ratios  for the 
disk plus star system, $\lambda_{\rm sys} \sim 4-12$.

Therefore, a current problem in protoplanetary disks is the effect of a strong poloidal magnetic field on the disk 
formation, structure, and emission. In this paper we study the vertical structure of magnetized
accretion disks taking into account viscous and resistive dissipation, as well
as stellar irradiation of the disk surface. We consider that the radial structure is given by the
analytic models of S07 but the formalism can be applied to other models 
of the disk radial structure. 
The paper is organized in the
following way: in \S 2 we briefly discuss the radial structure of the magnetized
disk models of S07; in \S 3  we discuss the equations of the vertical structure
that take into account internal heating and irradiation by the central star; in \S 4  we present 
the method of solution; in \S 5  we show the results of the models for accretion disks around 
three different YSOs with a mass-to-flux ratio $\lambda_{sys}=4$ and discuss the effect of the 
resistive dissipation and magnetic compression on the vertical structure; in \S 6 we discuss more weakly magnetized disks
with $\lambda_{sys}=12$ and compare with the former models; finally, in \S 7 we present the conclusions of this work. 

\section{S07 Radial Structure of Magnetized Accretion Disks }
\label{sec:Rad_struct}

During the process of gravitational collapse and disk formation, a fraction of
the magnetic flux from the parent core is dragged into the disk.
When the core accretion has ceased, the magnetized disk will
evolve subject to two diffusive processes: viscosity due
to turbulent and magnetic stresses that transfers angular momentum outside
and produces mass accretion toward the star, 
characterized by the coefficient $\nu$ $\{\cm^2 \, \s^{-1}\}$; and resistivity
due to microscopic collisions and the MRI which
allows matter to slip across field lines, characterized by the coefficient
$\eta$ $\{\cm^2 \, \s^{-1}\}$. S07 considered steady state models
where the dragging of field lines by accretion is balanced
by the outward field diffusion. In this case, 
$ \eta /\nu   \sim z_0/\varpi $, 
where $z_0$ is the vertical half  disk thickness and $\varpi$ is the radial cylindrical coordinate.

In near field freezing conditions, the accretion flow generates a
mean radial magnetic field from the mean vertical field
\footnote{The field lines are bent because the sources of the disk
magnetization are currents at infinity anchoring magnetic field
lines to the parent cloud.}.
This mean radial
field changes the radial
force balance and causes sub-Keplerian rotation of the gas.  If one
neglects the disk self-gravity and gas pressure, the force balance
equation is
\begin{equation}
\varpi \Omega^2 =  {GM_*\over \varpi^2} -{B_zB_\varpi^+\over 2\pi \Sigma_\varpi},
\label{centrifugal}
\end{equation}
where $\Omega$ is the rotation rate, $G$ is the gravitational constant,
$M_*$ is the stellar mass,  $B_z$ is the component of the magnetic field
threading vertically through the disk, and $B_\varpi^+$ is the steady state
radial component of the magnetic field just above the disk that 
responds to the radial accretion flow,
\be
B_\varpi^+ = -{ z_0 \nu \over \varpi \eta} \left( {\varpi \over \Omega} { d \Omega \over d \varpi}  \right) B_z.
\ee
We named the total radial mass
surface density $\Sigma_\varpi$ for further use in the vertical structure calculation.
 The rotation rate, given by
the solution of the above equation, is smaller than the Keplerian
value $\Omega_K = \left({GM_*/ \varpi^3}\right)^{1/2}$,
because of the extra support of the magnetic tension against
gravity. In particular, S07 studied the case when the rotation rate is a constant fraction of the Keplerian speed,
$ \Omega = f \Omega_K$, with the sub-Keplerian factor $f < 1$.

The stretching of the poloidal
field by differential rotation produces an azimuthal field in the
disk that, coupled with the radial field, exerts a mean stress and
torques the gas, allowing the disk viscous evolution.  S07 proposed a
functional form for the viscosity based on mixing length arguments,
\begin{equation}
\nu=D\frac{B_z^2 z_0}{2\pi \Sigma_\varpi \Omega},
\label{def_nu}
\end{equation}
where $D\le 1$ is a dimensionless viscosity coefficient
\footnote{ Note that eq. (2-3) is a prescription for the turbulent viscosity, only local
transport is considered.}.
This
coefficient acquires small values if there are substantial
``dead zones'', where the MRI would occur only on surface layers. S07 proposed that rapid transport
of mass and magnetic fluctuations across  strong mean field lines
can occur through the reconnection of small magnetic loops, twisted
and bent by the turbulent flow, this process being the source of
the disk viscous and resistive diffusivities.

With the viscosity $\nu$ given by eq. (\ref{def_nu}), and assuming a power-law disk aspect ratio 
$A(\varpi) = z_0/\varpi \propto  \varpi^n$, 
S07 constructed full steady state radial models of thin magnetized disks
around young stars.
 The radial structure of these magnetized disks is given by eqs. (63-69) of S07.
Four models are shown in their Table  2 calculated for an aspect ratio $A(\varpi) = A_0 (\varpi/100 \AU)^{1/4}$, assuming
 a mass-to-flux-ratio of the star plus disk system, $\lambda_{\rm sys} = 4$. 
 Given the stellar mass $M_*$, the disk accretion
rate $\dot M_d$, the system age $t_{\rm age}$, and the viscosity coefficient $D$, 
they calculated the sub-Keplerian
factor $f$, the disk radius $R_d$, and the disk mass $M_d$. As mentioned above, $f$ is constant,
independent of radius.

The S07 model assumes that the entire magnetic flux brought in by star formation
is contained in the disk, which gives the condition (S07 eq. [60])
\begin{equation}
1-f^2 = {0.5444\over \lambda_{\rm sys}^2 } \left({M_* \over M_d} \right).
\label{eq:ft}
\end{equation}  
For a closed star plus disk system in which infall has ceased, the
mass-to-flux ratio $\lambda_{\rm sys}$ remains constant. Since
disk accretion decreases the disk mass $M_d$  relative to stellar
mass $M_*$, the departure from Keplerian rotation, $(1-f^2)$, must
grow with time.  This happens because viscosity drains mass from
the disk onto the star, while resistivity can only cause the
redistribution of flux within the disk but cannot change the total
flux. Thus, $f$ decreases with time, i.e., the disk becomes more
sub-Keplerian and magnetized with time. 
{ S07 showed that, even in the case of a strong poloidal field, their disk models
fulfill the condition to develop the MRI, i.e., that the gas to magnetic pressure $\beta > 1$. 
Nevertheless, the stability of these disks to the MRI has not been studied. In the rest of the paper 
we will assume that the MRI operates and we will calculate the heating due to the associated viscosity. 
This problem is further discussed in \S 6. }

Finally, in steady state, the energy flux has to carry the sum of the energies generated by viscous and resistive dissipation 
inside the disk.  At each radius, the viscous dissipation 
 rate per unit area gives the flux $\{ {\rm  erg \, s^{-1} cm^{-2}} \}$ 
 \be
F_{v, \varpi}= \nu \Sigma_\varpi\left( \varpi \frac{d \Omega}{d \varpi} \right)^2  = \frac{9}{4} \nu  f^2 
\Omega_K^2 \Sigma_\varpi.
\label{eq:Vis}
\ee
From eqs. (11), (12) and (33) of S07, the flux due to resistive dissipation is 
\be
F_{r,\varpi}=\left( \frac{c B_\varpi^+ }{2 \pi}  \right) \left(-\frac{u}{c} B_z \right) =  
{\eta \over z_0} {(B_\varpi^+)^2 \over 2 \pi } =
 \frac{3}{2} \frac{\nu}{\varpi} \frac{ B_\varpi^+ B_z}{2 \pi},
\label{eq:Res}
\ee
where $u$ is the radial accretion velocity and $c$ is the speed of light.
In steady state, the internal energy flux 
\be 
F_{vr,\varpi} = F_{v,\varpi}+F_{r,\varpi}
\label{eq:internal_flux}
\ee 
carries away the energy generated by viscous and resistive dissipation 
inside the disk. 

In the next section we discuss the equations of the
vertical structure of magnetized disks including the surface irradiation by a central source.

\section{Vertical Structure of Magnetized Viscous Disks}

The equations for the vertical structure are derived
following D'Alessio et al. (1998). We use the radial structure of the S07 models discussed above for a
 a thin cold accretion disk with negligible mass
compared to the central star. 
The magnetized disk is subject to both viscous and resistive heating and is also irradiated by the central star.
The stellar radiation flux, $F_{irr}$ will penetrate the disk atmosphere down to the irradiation surface $z_{irr}$ 
where the optical depth is 1.  For simplicity, will assume that the sub-Keplerian factor $f$ is constant with height,
since the vertical variation of the azimuthal velocity is small in thin disks (e.g., Li 1995). {We will also assume
that the viscosity and resistivity coefficients are independent of $z$ (see discussion in \S 6).}

We use as the independent variable the midplane mass surface density defined as
\be
\Sigma (\varpi,z)  = \int_0^z \rho(\varpi,z) dz,
\label{Sigmamid}
\ee
 where $\rho(\varpi, z)$ is the density and $z$ is the height, such that 
 the total radial mass surface density, integrated from the disk surface $z_\infty$ below and above the plane, is
\be
\Sigma_\varpi = \int_{-z_\infty}^{z_\infty} \rho dz = 2 \, \Sigma(\varpi, z_\infty),
\ee
where the surface $z_\infty$ is defined as the height where
the disk pressure is equal to an external pressure $P_\infty$.

Following Calvet et al. (1991), the thermal source function $B(T)$ is given by the superposition
of the source function of a non-irradiated viscous and resistive disk with temperature $T_{vr}$, 
plus the source function of an irradiated passive disk without internal energy sources that reprocess the
stellar radiation and has a 
temperature $T_{rp}$. Appendix A discusses the 
first and second moments of the 
transport equations for the viscous resistive flux $F_{vr}$ and mean intensity $J_{vr}$ of the
non-irradiated viscous resistive disk, eqs. (\ref{Tvr}) - (\ref{dJvr}), and 
the reprocessed flux $F_{rp}$ and mean intensity $J_{rp}$ of the passive irradiated disk,
eqs. (\ref{Trp}) - (\ref{dJrp}). Because 
 the transport equations are linear in the fluxes, they can be added such that 
the local disk temperature $T$ and mean intensity $J$ $\{\rm erg\, cm^{-2} s^{-1} sr^{-1}\}$ are given by
\be
T^4 = T_{vr}^4 + T_{rp}^4, \quad {\rm and } \quad J=J_{vr}+J_{rp},
\label{eq:T4}
\ee
and the equation of energy transport in the disk becomes
\be
\frac{\sigma T^4}{\pi} =J +\frac{1 }{ 4 \pi \kappa_P}   \frac {d \left(
F_{vr}
+ F_{rp} \right)} {d \Sigma},
\label{Temp}
\ee
where $\sigma$ is the Stephen-Boltzmann constant, and
 $\kappa_P$ is the Planck mean opacity. With the total disk flux $F_{vr}$ + $F_{rp}$ given by
 eqs. (\ref{eq:fluxvr}) and (\ref{Frp}), 
 the energy equation becomes simply an algebraic equation for the temperature.
 The equation for total mean intensity $J$  becomes
\be
\frac{d J}{d\Sigma} = - \frac{3}{4 \pi} \chi_R \left( F_{vr} + F_{rp}\right),
\label{dJmean}
\ee
where $\chi_R$ is the Rosseland mean opacity.

The equation of hydrostatic equilibrium in the vertical direction is
\be
\frac{d P}{d \Sigma} = -\frac{GM_*}{\varpi^3} \frac{z}{  \left[ 1 + \left( \frac{z}{\varpi}\right)^2 \right]^{3/2} } 
 -{d P_{rad}  \over d \Sigma}-\frac{1}{8 \pi}\frac{d B_\varpi^2}{d \Sigma},
\label{HE}
\ee
where $P$ is the gas pressure, $P_{rad}$ is the radiation pressure, and the last term is the magnetic pressure, where 
the azimuthal component of the field is neglected.
With the same assumptions as in eq. (\ref{centrifugal}), the radial force balance at each height gives
\be
\frac{B_z}{4 \pi} \frac{d B_\varpi}{d \Sigma} = \varpi \Omega_K^2 \left[ 1 - f^2 \right],
\ee
 where the net radial force associated with the departure from Keplerian rotation is balanced by the magnetic tension due to 
the bending of the poloidal field lines. Since the right-hand side of this equation is a function of the radius only, the radial component of the 
field, $B_\varpi$, is a linear function of $\Sigma$,
$B_\varpi = 2 B_\varpi^+{(\Sigma}/{\Sigma_\varpi})$.
The radiation pressure force due to 
the viscous resistive, reprocessed, and scattered
 fluxes in the positive $z$ direction, minus the 
radiation pressure of the stellar flux $F_{irr}$ (eq. \ref{eq:flux-2}) that enters the disk at the upper surface, is
\begin{eqnarray}
- {d P_{rad}  \over d \Sigma}
&  = & {\chi_R \over c} \left[ F_{vr} + F_{rp} \right] + { \chi_P^s \over c} \left[ F_s - 
            F_{irr} \exp^{-\tau_s/\mu_0} \right], \\ 
            & = & {\chi_R \over c} F_{vr}  +{ \left( \chi_R - \chi_P^s \right) \over c} F_{rp},
\end{eqnarray}
where $F_s$ is the scattered flux, $\tau_s$ is the optical depth  
to the stellar radiation, $\chi_P^s$ is the Planck mean opacity at the stellar temperature (see eq. \ref{taus}),
 and the last equality comes from the zero flux condition in eq. (\ref{Frp}).

For an ideal gas, the pressure is
\be
P = { \rho k T \over \mu m_H},
\label{Pideal}
\ee
where $k$ is the Boltzmann constant, $\mu$ is the mean molecular weight, and $m_H$ is the hydrogen mass.
From the definition $d \Sigma = \rho dz$,  one obtains a differential equation for the height 
\be
\frac{d z}{d \Sigma }= \frac{kT}{\mu m_H P}.
\label{scaleheight}
\ee

Finally, the vertical structure is given by one algebraic equation for the temperature (\ref{Temp}) and 
three differential equations (\ref{dJmean}), (\ref{HE}), and (\ref{scaleheight}), with BCs 
 imposed at the upper disk boundary where
\be
z(\Sigma_\varpi/2)=z_\infty, \quad P(\Sigma_\varpi/2) = P_\infty , 
\quad {\rm and}  \quad J(\Sigma_\varpi/2)=J_\infty, 
\ee
where the mean intensity and flux at the disk surface $z_\infty$ is obtained from
eqs. (\ref{eq:BCvr}) and (\ref{eq:BCrp}),
\be
J_\infty = 
{\sqrt{3} \over 4 \pi}  F_\infty,  \quad {\rm and} \quad 
F_\infty =  {F_{vr,\varpi} \over  2} + F_{irr} a_s( 1 + C_1 + C_2),
\label{eq:Finfty}
\ee
where  the fractional absorption is $a_s$ is defined by eq. (\ref{eq:frac_abs}), and the 
constants $C_1$, $C_2$ are given by eqs. (\ref{C12}).

This set of equations is then 
solved for  the temperature $T$, the mean intensity $J$, the pressure $P$, and the height $z$ as functions of the
midplane surface density $\Sigma$.
In this way, the disk vertical structure is obtained for each radius $\varpi$.  

In the next section we present the non dimensional equations that we solve.


\subsection{Non Dimensional Equations}

We define the non dimensional surface density and radius
\be
s=\frac{\Sigma}{\Sigma_\varpi} ;  \quad {\rm and} \quad
 r=\frac{\varpi}{z_\infty}. \nonumber
 \ee
At each radius $r$, the non dimensional height, temperature, mean intensity, 
and pressure are defined as
 \be
 \zeta=\frac{z}{z_\infty};  \quad  t=\frac{T}{T_\infty}; \quad  j=\frac{J}{J_\infty}; \quad {\rm and} \quad p=\frac{P}{\Sigma_\varpi \Omega_K^2 z_\infty};  
 \nonumber
 \ee
 where the constant  $T_\infty=\left({F_\infty}/{\sigma}\right)^{1/4}$, with $F_\infty$ given by eq. (\ref{eq:Finfty}).
 The non dimensional vertical energy fluxes are 
 \be
 f_{vr}=\frac{F_{vr}}{F_\infty}, 
 \quad  f_{rp}=\frac{F_{rp}}{F_\infty}, \quad {\rm and} \quad f_{irr} = \frac{F_{irr}}{F_\infty} , \nonumber
 \ee
 and the non dimensional total viscous resistive energy flux is 
 $f_{vr,r} = F_{vr,\varpi}/F_\infty$.

For uniform dissipation $f_{vr}= f_{vr,r} \, s$, the transport equations are:
the energy equation
\be
4 t^4 = \sqrt{3}j+ 
\frac{1}{  \Sigma_\varpi \kappa_P }
\left( f_{vr,r} + {d{ f_{rp}}  \over ds} \right), 
\label{temp}
\ee
where 
\be
{d f_{rp} \over ds}  = \Sigma_\varpi \chi_P^s  f_{irr}  a_s  \left( \frac{(1+C_1)}{\mu_0} e^{-\tau_s/\mu_0} + \beta_s C_2 e^{-\beta \tau_s} \right),
\ee
$\beta_s=\sqrt{3 a_s}$, and the opacity to the stellar radiation is 
 \be
 \tau_s = \Sigma_\varpi  \int_s^{1/2} \chi_P^s ds ;
 \ee
and the  mean intensity equation 
\be
\frac{dj}{ds} = - \sqrt{3} \Sigma_\varpi \chi_R \left(f_{vr,r}\, s+ f_{rp} \right).
\label{jmean}
\ee
The  hydrostatic equilibrium equation is
\be
\frac{d p }{d s} = -\frac{\zeta}{\left( 1+\left(\frac{\zeta}{r}\right)^2 \right)^{3/2} } - C_3 {d p_{rad} \over ds} -C_4 s,
\label{HEnd}
\ee
where the radiation pressure force is
\be
-{d p_{rad} \over ds} = \Sigma_\varpi \chi_R f_{vr,r}\, s +\Sigma_\varpi \left( \chi_R -  \chi_P^s \right)  f_{rp} \, .
\ee
Finally, the height equation is 
\be
\frac{d \zeta}{ d s} = C_5 \frac{t}{p \mu}.
\label{heightnd}
\ee
In the above equation, the non dimensional constants are
 \be
 C_3={F_\infty \over c  \Sigma_\varpi \Omega_K^2 z_\infty}, \quad
 C_4= \frac{(B_\varpi^+)^2}{\pi \Sigma_\varpi  \Omega_K^2 z_\infty } ,
 \quad
  \quad {\rm and} \quad  C_5={k T_\infty  \over m_H \Omega_K^2 z_\infty^2}.
 \label{eq:C3C4}
 \ee 
 
 Given the irradiation flux at the disk surface, $f_{irr}$,
 eqs. (\ref{temp}) - (\ref{heightnd}) can be solved for the
 non dimensional  temperature $t(s)$, mean intensity $j(s)$, pressure $p(s)$ and height $\zeta(s)$. 
 The  BCs at the upper boundary $s=1/2$ are given by
 \be
j(1/2)=1 ; \quad p(1/2) = P_\infty/(\Sigma_\varpi \Omega_K^2 z_\infty); \quad {\rm and} \quad \zeta(1/2)=1.
\label{BCs}
\ee
There is an extra condition at the midplane 
\be
\zeta(0)=0,
\ee
that is satisfied for the appropriate eigenvalue $z_\infty$
which determines the constants $C_3$, $C_4$, $C_5$, and the boundary value $p(1/2)$.
This eigenvalue has to be found by iteration process, as discussed in below.
 
Finally, this set of equations is solved for each non dimensional radius $r$ to build
 the full disk vertical structure. 
 
 \section{ Method of Solution}
 \label{sec:MofS}

We use eqs. (63), (64) and  (67) of S07 to obtain the disk radial structure:
the total surface density $\Sigma_\varpi$, the vertical and radial magnetic 
fields $B_z(\varpi)$ and $B_{\varpi}^+(\varpi)$, and the viscosity $\nu(\varpi)$. From eq. (\ref{eq:internal_flux}), 
 one obtains the total viscous resistive flux at each radius $F_{vr,\varpi}$,
 one of the main ingredients of the model.

One first solves for the vertical structure of a
 non irradiated viscous resistive disk with $f_{irr} = 0$.  One guesses
 the disk surface $z_\infty$ that is an eigenvalue, and 
 calculates the constants $C_3$, $C_4$, $C_5$ (eq. \ref{eq:C3C4}), and the boundary value $p(1/2)$  in eq. (\ref{BCs}). 
 The differential equations eqs. (\ref{temp}) - (\ref{heightnd}) are integrated from the surface $s = 1/2$ toward the midplane $s=0$.
  In general, as a result of the 
 integration, $\zeta(0) \neq 0$. One modifies $z_\infty$ until the eigenvalue is found such that $\zeta(0) = 0$ 
 This eigenvalue is used as a first guess for the irradiated disk structure.
 
Given the disk vertical structure with $f_{irr} = 0$, one calculates the irradiation surface $z_{irr}$,
 the angle of irradiation $\mu_0$, and the intercepted flux $f_{irr}$ on this surface, as shown in Appendix B.
Then, one solves the equations for the vertical disk structure with the same iteration process as discussed above, 
until $\zeta(0) =0$, to obtain a new
 structure and disk surface $z_\infty$ for all the disk. Given this disk structure, one 
obtains a new irradiation surface, angle $\mu_0$, and irradiation flux $f_{irr}$ at each radius,
to calculate a new vertical structure. One repeats this procedure until the surface $z_\infty$ converges.

For the dust  composition we adopt a mixture of silicates, organics, and ice
with a mass fractional abundance with respect to gas 
 $\zeta_{sil} = 3.4 \times 10^{-3}$, 
$\zeta_{org} = 4.1 \times 10^{-3}$, and $\zeta_{ice} = 5.6 \times 10^{-3}$,
with bulk densities $\rho_{sil} = 3.3 \, {\rm g \, cm}^{-3}$, $\rho_{org} = 1.5 \, {\rm g \, cm}^{-3}$,
and $\rho_{ice} = 0.92 \, {\rm g \, cm}^{-3}$ 
(e.g., Pollack et al. 1994).
The dust particles have a power-law size distribution, $n(a) \sim a ^{p}$, with exponent $p=3.5$,  a minimum 
grain size $a_{min} = 0.005 \, \mu$m, and maximum grain size $a_{max} = 1$ mm, and are considered compact. The 
value of $a_{max}$ is consistent with evidence of grain growth in millimeter spectral energy distributions of protoplanetary 
disks around YSOs, see, e.g., Ricci, Testi \& Natta  (2010). 
With this dust mixture, we obtain the Planck and Rosseland mean opacities at both the dust and
the stellar temperatures.
We assume well mixed dust and gas, and leave for a future study the 
effect of dust settling (e.g.,  D'Alessio et al. 2006;
Boehler et al. 2013; Guilloteau et al. 2011; Gr{\"a}fe et al. 2013) and 
dust radial migration  (e.g., Brauer et al. 2007; Birnstiel et al. 2010; P\'erez et al. 2012; 2015).

 \section{ Results}
 
 We obtain the vertical structure of a low mass protostar disk (LMP), a FU Ori disk, and a T Tauri disk around a
 central star with a mass $M_* = 0.5 \msun$.  We consider the standard disk models of S07
 with a mass-to-flux ratio $\lambda_{sys}=4$ shown in their Table 2. This table gives 
the values of the mass accretion rate $\dot M_d$, the viscosity coefficient $D$,  
the disk mass $M_d$, and sub-Keplerian parameter $f$.
To irradiate the disk, we assume a central source
 characterized by a stellar radius $R_*$ and an effective  temperature 
 $T_{eff} = (L_c/ 4 \pi R_*^2 \sigma )^{1/4}$, such 
that it produces the total central source luminosity 
$L_c$ (accretion plus stellar luminosities). The FU Ori case has the highest accretion
rate and thus, the highest value of $T_{eff}$. These parameters are sumarized in Table \ref{table:1}.

To compare with observations, at each radius we define the observed surface density, $\Sigma_{obs}$,
 measured from the disk surface $z_\infty$ towards the midplane, such that
\be
\Sigma_{\rm obs}(z) = {\Sigma_\varpi \over 2} - \Sigma(z).
\ee

\begin{figure} 
\centering
\includegraphics[angle=0,width=0.8\textwidth]{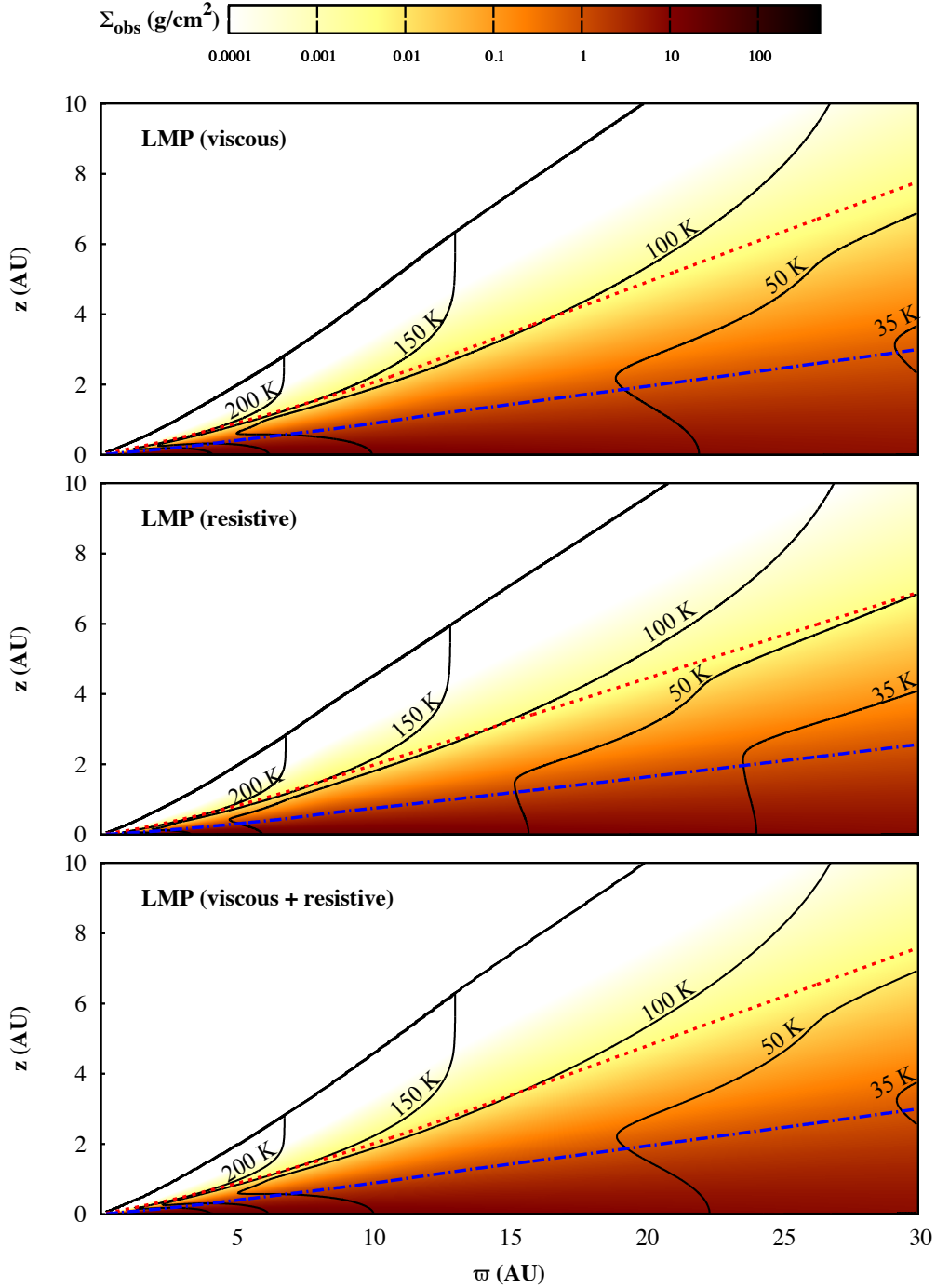}
\caption{
Vertical structure of the low mass protostar (LMP) disk 
with a mass-to-flux ratio $\lambda_{sys}=4$: the contours show the temperature and the color scale represents the
surface density measured from the disk surface $\Sigma_{obs}$. The dashed red line shows the irradiation surface $z_{irr}$.
The dot-dashed blue line shows the disk mass surface $z_{90}$.
The model in the upper panel includes only viscous heating, in the middle panel 
only resistive heating, and the model in the lower panel includes both viscous and resistive heating. 
}
\label{fig:LMP}
\end{figure}

\begin{figure} 
\centering
\includegraphics[angle=0,width=0.8\textwidth]{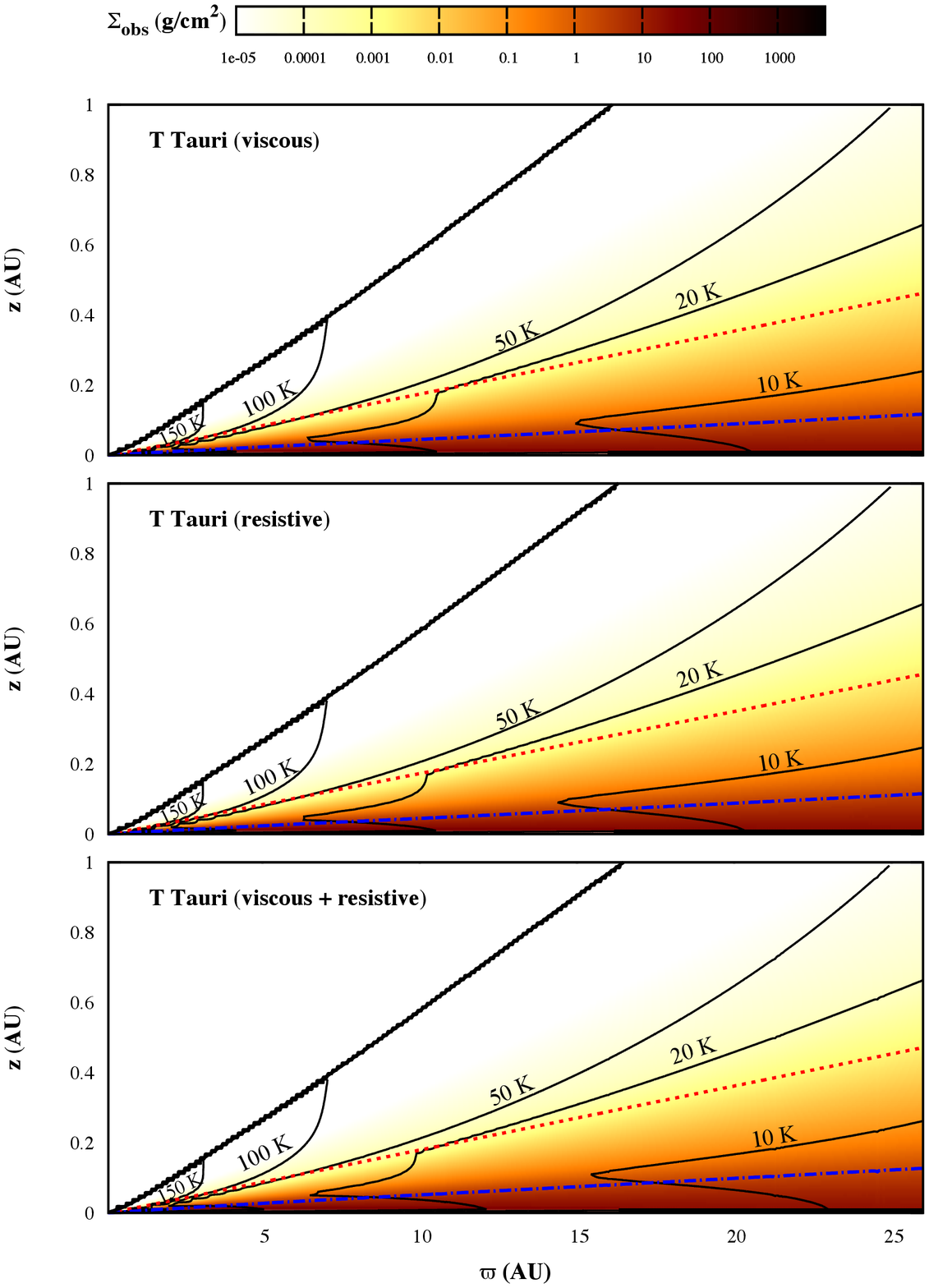}
\caption{
Vertical structure of the T Tauri disk with a mass-to-flux ratio $\lambda_{sys}=4$; same description as in Figure \ref{fig:LMP}.
}
\label{fig:TT}
\end{figure}

\begin{figure} 
\centering
\includegraphics[angle=0,width=0.8\textwidth]{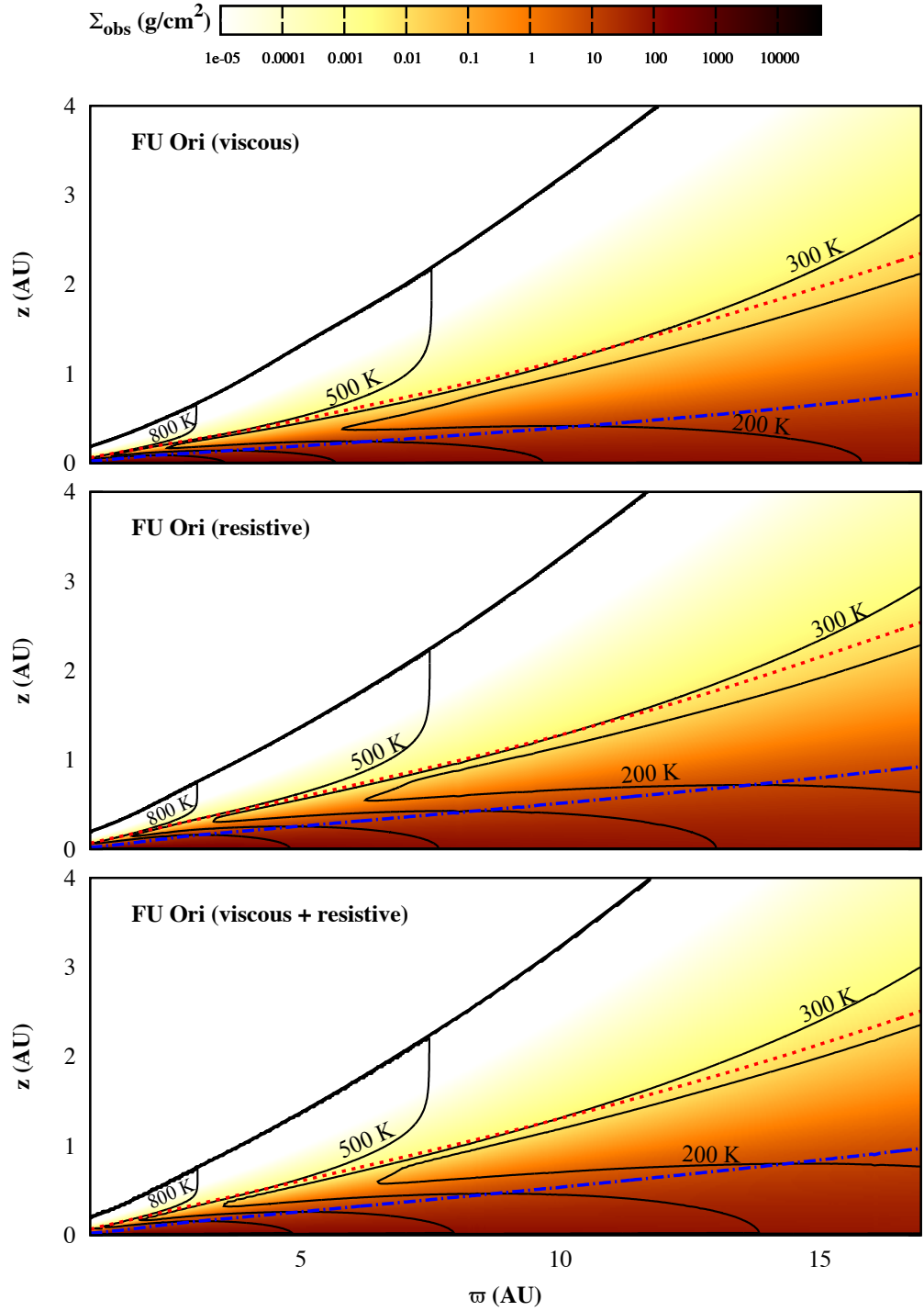}
\caption{
Vertical structure of the FU Ori disk with a mass-to-flux ratio $\lambda_{sys}=4$; same description as in Figure \ref{fig:LMP}.
}
\label{fig:FUOri}
\end{figure}

Figures (\ref{fig:LMP}),  (\ref{fig:TT})  and (\ref{fig:FUOri}), show the vertical structure of the LMP,  T Tauri, and
FU Ori disk,  respectively. 
In each figure, the upper panel includes only the viscous heating $F_{v,\varpi}$ in eq. (\ref{eq:Vis});
the middle panels considers only the 
resistive heating $F_{r,\varpi}$ in eq. (\ref{eq:Res}); and
 the bottom panel shows the vertical structure of the disk taking into account both 
viscous and resistive heating $F_{vr,\varpi}$.
The dashed red line shows the irradiation surface $z_{irr}$. 
To quantify the location of the disk mass, 
we define the disk mass surface, $\pm z_{90}$, 
as the surface that contains 90\% of the total  surface density, $\Sigma_\varpi$, above and below the midplane,  
i.e., $ \pm z_{90} = \pm z(0.45 \Sigma_\varpi)$.
This surface is shown as the dot-dashed blue line in every panel. To calculate the radial structure 
we assign this surface to the vertical half  disk thickness $z_0 $ in the S07 model, i.e., 
the aspect ratio for the radial structure in eqs. (63), (64) and  (67) of S07
 is given by $A(\varpi) = A_{\lambda_{sys}} (\varpi / 100 AU)^{1/4}$, 
where $A_{\lambda_{sys}}\equiv z_{90}(100 AU) /{100 \rm AU}$. We iterate the models until this surface is fixed and obtained
$A_{4} = 0.156, 0.013$, and $0.102$, for the LMP,  T Tauri, and FU Ori disk, respectively.

The irradiation flux  heats the hot upper atmosphere, while the midplane can be dominated by the viscous and
resistive heating.
In the S07 models, the ratio of resistive to viscous dissipation is given by the factor $2(1-f^2)/3 f^2$
 (see their eqs. [32] and [33]). For the case of the LMP disk, with the sub-Keplerian parameter
 $f_{4}= 0.957$ the ratio is $F_{r,\varpi}/F_{v,\varpi} \sim 0.06$, i.e.,
 the resistive heating is negligible compared to the viscous heating. 
Thus, the viscous model (upper panel) in Figure (\ref{fig:LMP}) is hotter in the midplane 
than the resistive model (middle panel), and viscous heating dominates the thermal
structure in the bottom panel.
For the case of the T Tauri disk with $f_{4}= 0.658$, the ratio is $F_{r,\varpi}/F_{v,\varpi} \sim 0.87$,
i.e., both fluxes have similar contributions. 
For the case of the FU Ori disk $f_{4}= 0.386$ and the ratio is $F_{r,\varpi}/ F_{v,\varpi} \sim  3.8 $.
Thus, the resistive FU Ori model in Figure (\ref{fig:FUOri})  is hotter in the midplane than
the viscous model, and resistive heating dominates the thermal structure in the bottom panel.
Also, the LMP and the FU Ori disks are hotter and thicker than the T Tauri disk.  The T Tauri disk is 
highly compressed by the magnetic pressure and is very flat and cold.

\begin{figure}
\centering
  \includegraphics[angle=0,width=0.78\textwidth]{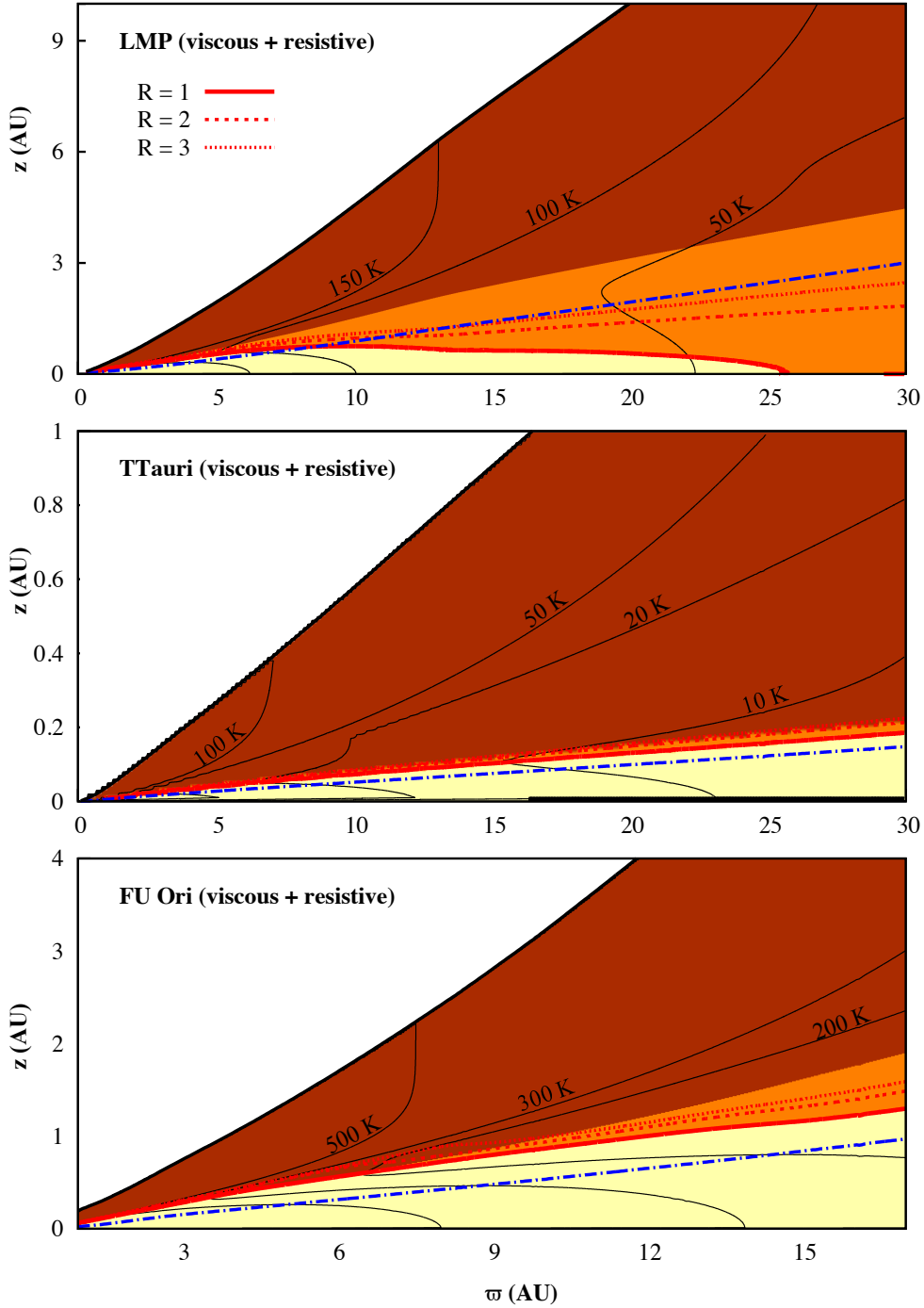}
  \caption{Dominant heating sources inside the LMP, T Tauri and FU Ori disks with a mass-to-flux ratio $\lambda_{sys}=4$:
  the yellow zone is the active zone where the viscous resistive 
heating dominates;  in the orange zone 
the reprocessed mean intensity $J_{rp}$ dominates; 
and in the red zone the irradiation flux is absorbed.
The dot-dashed blue line in 
each panel is surface $z_{90}$. 
The solid red lines correspond to the ratio ${\cal R} \equiv (T_{rp}/T_{vr})^4 = 1$, 
the dashed red lines correspond to  ${\cal R}= 2$, and the dotted red lines correspond 
${\cal R} = 3$. }
  \label{fig:act-pass}
\end{figure}

Figure \ref{fig:act-pass} shows the dominant heating sources inside the LMP, T Tauri, and  FU Ori  disks
in the bottom panels of Figures (\ref{fig:LMP}) - (\ref{fig:FUOri}).  The dot-dashed blue line in each panel is $z_{90}$.
The yellow zone at the disk midplane is the so-called
active zone where the viscous resistive heating determines the disk temperature, i.e., $T_{vr} \ge T_{rp}$,
see eq. (\ref{Temp}). In this zone the temperature decreases with height.
The red zone indicates the region where
the irradiation flux is absorbed, this heating decreases with depth measured from the disk surface, as $\tau_s$ increases
\footnote{The gradient is negative with respect to $\Sigma_{obs}$, i.e., 
$F_{rp} \rightarrow 0$ inside the disk.}.  Both effects produce a temperature inversion at the base of the
hot atmosphere. In the orange zone  the mean intensity $J_{rp}$ dominates the heating.
In this region, when $F_{rp} \sim  0$, $J_{rp}$ is constant (eq. \ref{dJrp}); thus, at large radii 
midplane passive regions tend to be vertically isothermal. 
The solid red lines
corresponds to the ratio ${\cal R}  \equiv (T_{rp}/T_{vr})^4 = 1$, 
the dashed red line corresponds to  ${\cal R} = 2$, and the dotted red line corresponds 
${\cal R} = 3$. 
The viscous resistive heating still contributes to the heating inside ${\cal R} \sim 3$
and deviates the temperature from the vertically isothermal regime.

The base of the hot atmosphere is easily identified by the location of a sharp
transition in the temperature isocontours (elbow), due to the temperature inversion discussed above.
The figures show that  the 
hot atmosphere extends close to the mass surface $z_{90}$ such that, it contains $\leq $ 10\% of the 
total surface density $\Sigma_\varpi$, 
with half of the mass on each atmosphere above and below the midplane.
One expects the extension of the hot atmosphere to change in models that consider dust settling where
only small grains ($ a < 10 \, \mu$m) survive in the upper disk layers. 
On the one hand, small grains will absorbe the irradiation flux more
efficiently than in the well mixed case we consider here, on the other hand, one expects a lower opacity in 
the atmosphere because the dust mass has settled to the midplane (D'Alessio et al. 2006). 
This will be a subject of a future study.

\begin{figure}
\centering
  \includegraphics[angle=0,width=0.8\textwidth]{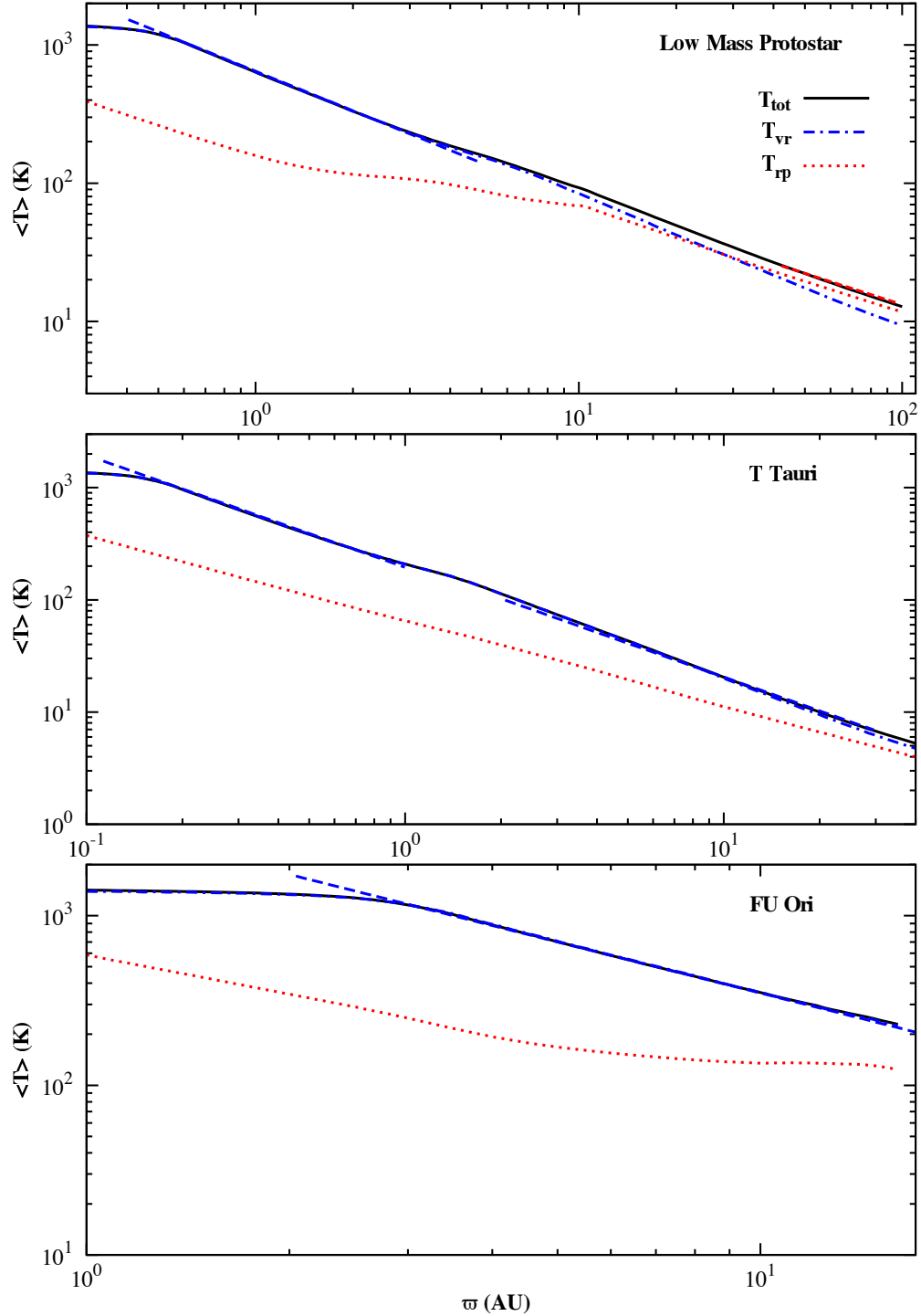}
  \caption{Mass weighted temperature for LMP, T Tauri, and FU Ori disks as functions of the radius ($\lambda_{sys}=4$). The 
 dotted red lines  correspond to the reprocessed temperature $<T_{rp}>$ due to the external heating by
the central source. The dot-dashed blue lines correspond to the viscous resistive temperature $<T_{vr}>$ due to
 internal heating. 
The temperature $<T>$  in solid black line takes into account both external  and internal heating.
The dashed blue lines in each panel have a slope $s=-1$. In the upper panel of the LMP disk,  the dashed red  line has
a slope $s=-3/4$.}
  \label{fig:temp}
\end{figure}

Figure \ref{fig:temp} shows the mass weighted disk temperature as function of radius, 
 for the LMP,  T Tauri,  and FU Ori disks, indicated in each panel.
The dotted red line in each panel corresponds to the reprocessed temperature  $<T_{rp}>$ due to the external heating by
the central source. The dot-dashed blue line in each panel corresponds to the viscous and resistive temperature $<T_{vr}>$
due to the internal heating. 
The solid black line indicates the total temperature $<T>$,  that takes into account both external  and internal heating
(eq. \ref{eq:T4}). 
The intersection between the $<T_{vr}>$ and $<T_{rp}>$ curves at $R_{\rm active}$
gives an estimate of the border between the active region, where internal heating dominates,
and passive region dominated by the external heating. The size of the active region corresponds to 
the radial size of the region with ${\cal R } =1$ in Figure \ref{fig:act-pass}.
In the upper panel of the LMP disk, one can see that $R_{\rm active} \sim 25$ AU. In contrast, the midplane of both the 
 T Tauri and FU Ori disks is active.  
The  mass weighted temperature in the active regions follows a power law 
$<T > \sim \varpi^{-1}$ as  shown by the dashed blue lines in each panel. 
As shown by the dashed red line in the top panel, the temperature in the passive
 region of the LMP disk has a shallower slope $<T> \sim \varpi^{-3/4}$, which is the expected value for flat passive
 disks with $z_{irr} \propto \varpi$ 
 (e.g., Friedjung 1985).  From the model emission, one could mimic the observational procedure
 used to obtain the radial dependence of the disk temperature to compare
 with observations at a given set of wavelengths (e.g, Guilloteau et al. 2011).
 
 \section{Discussion}

In the previous section we found the vertical structure of disk models with a mass-to-flux ratio $\lambda_{\rm sys}=4$,
which correspond to the standard models discussed by S07.
These disks are strongly magnetized and are compressed by both gravity and magnetic pressure. In the case
of the T Tauri disk, the magnetic pressure dominates the compression and the disk is very thin.
Furthermore, the disk is cold because it intercepts little stellar irradiation.
For the latter reason, as shown above, the disk has a large active region.
In fact, the T Tauri disk has an aspect ratio at 100 AU,
 $A_{4} = 0.013$. Nevertheless, observations of disks around T Tauri stars, assuming isothermal vertical structures,
 infer scale heights at 100 AU, $H \sim 4 - 20 $ AU (e.g., Andrews et al. 2009; Pinte et al. 2008, Gr{\"a}fe et al. 2013).
 These values of $H$ would correspond  to aspect ratios 
$z_{90}/100 AU \sim \sqrt{2} H / 100 AU \sim 0.06 - 0.28$, larger than the T Tauri disk model.

\begin{figure}
\centering
  \includegraphics[angle=0,width=0.8\textwidth]{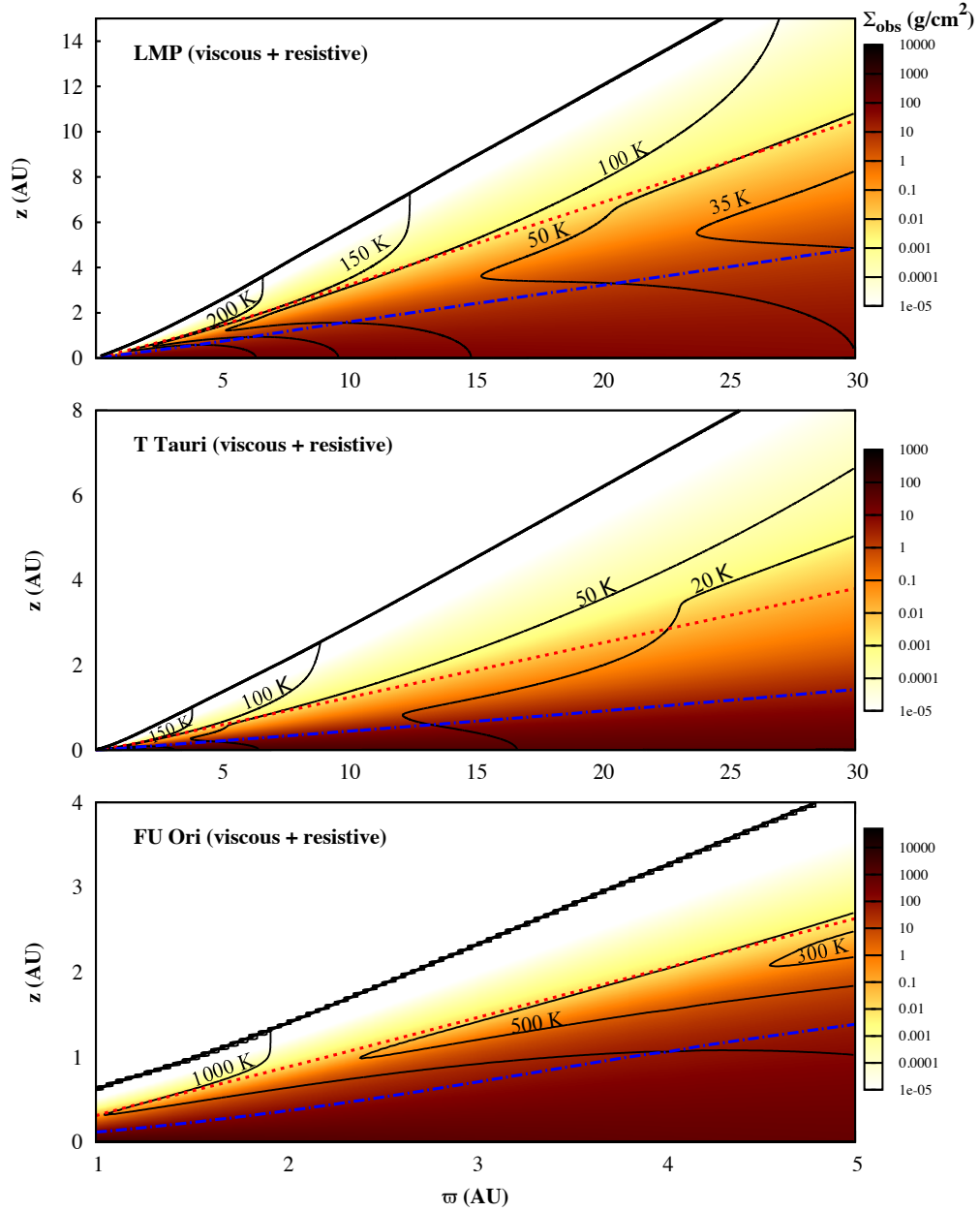}
  \caption{Vertical structure of a low mass protostar LMP, T Tauri, and FU Ori disk, with a mass-to-flux ratio $\lambda_{sys} = 12$. 
  The contours show the temperature and the color scale represents the
surface density measured from the disk surface $\Sigma_{obs}$. The dashed red line shows the irradiation surface $z_{irr}$.
The dot-dashed blue line shows the disk mass surface $z_{90}$.
The models include both viscous and resistive heating. }
  \label{fig:lambda12}
\end{figure}

For this reason it is relevant to consider disk models with a weaker magnetic field. 
 The disk properties change for different values of $\lambda_{sys}$:
for the same stellar mass $M_*$ and disk mass $M_d$ but larger  values of 
$\lambda_{\rm sys}$ (lower magnetization), the sub-Keplerian factor $f$ given by eq. (\ref{eq:ft}) is closer to 1,
decreasing the contribution of resistive heating. The different value of $f$ changes the disk radial
surface density $\Sigma_\varpi$ and disk size (for the same disk mass).  
We calculate disks models with $\lambda_{\rm sys} =12$, for the parameters given in
Table  \ref{table:1} (except $f$).
Figure \ref{fig:lambda12} shows the vertical structure of the LMP, T Tauri, and FU Ori disk.
These models
are  warmer and thicker than the $\lambda_{\rm sys} = 4$ models. 

Table  \ref{table:2} shows the values of the sub-Keplerian parameter $f_{\lambda_{\rm sys}}$ and the aspect ratio 
$A_{\lambda_{\rm sys}}$ at 100 AU  for the
models with mass-to-flux ratio $\lambda_{\rm sys} = 4$ and $\lambda_{\rm sys} = 12$. The aspect ratio is
given by $A= A_{\lambda_{\rm sys}} (\varpi/100 \, {\rm AU})^{-1/4}$.
For the models with $\lambda_{\rm sys} = 12$, 
the resistive dissipation is only 1\%, 5\% and 7\%  of the viscous heating for the LMP, T Tauri, and FU Ori disk, respectively.
The weakly magnetized T Tauri disk  has
 an aspect ratio $A_{12} = 0.109$, more consistent values inferred from observations. 
 
\begin{figure}
\centering
  \includegraphics[angle=0,width=0.78\textwidth]{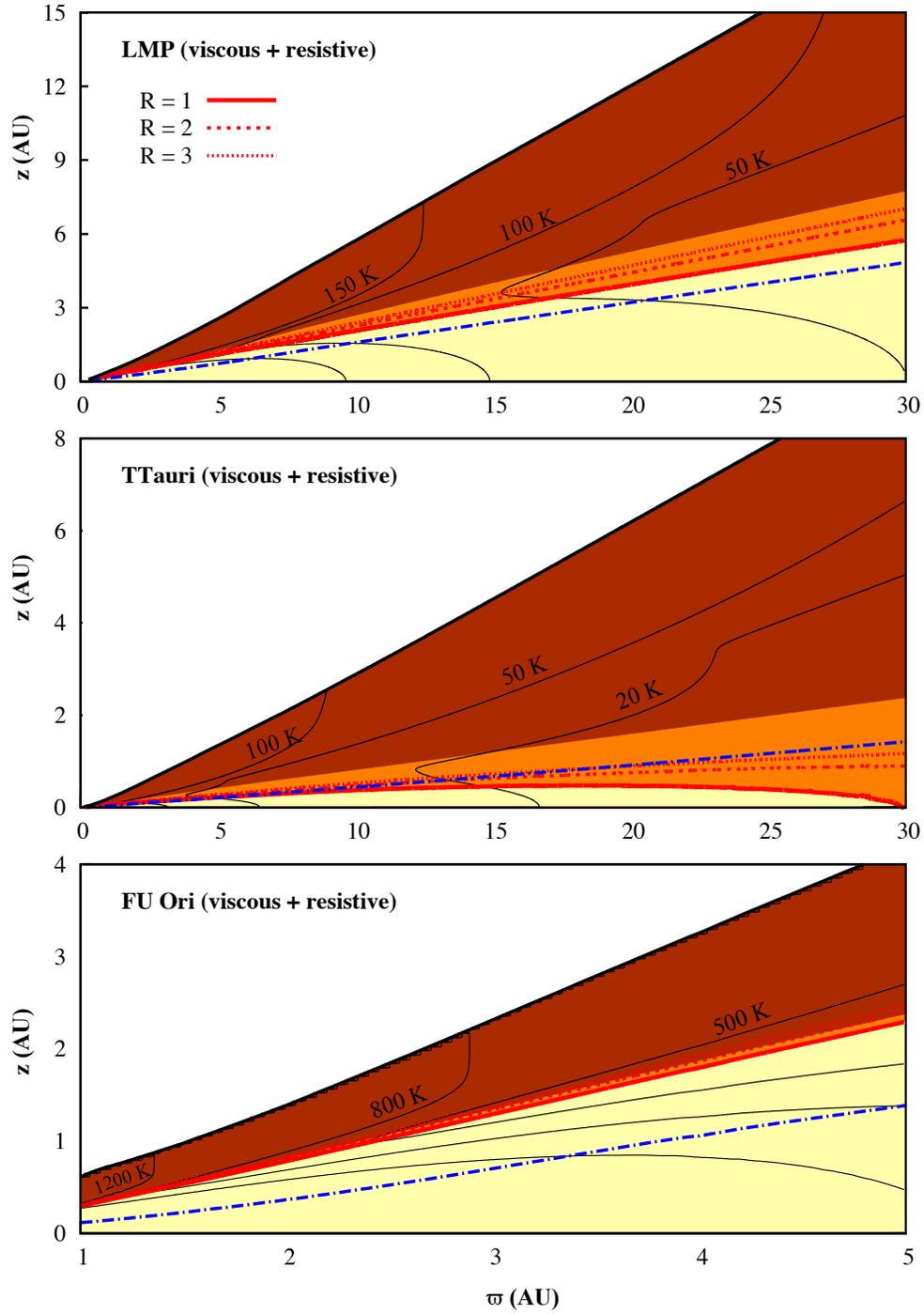}
  \caption{Dominant heating sources inside the LMP, T Tauri and FU Ori disks with $\lambda_{\rm sys} = 12$;
 same description as in Figure \ref{fig:act-pass}.}
  \label{fig:act-pass12}
\end{figure}

In addition, Table  \ref{table:2} shows the values of the surface density at 100 AU for models with different
mass-to-flux ratios $\Sigma_{\lambda_{sys}}$ where the radial 
surface density is given by $\Sigma = \Sigma_{\lambda_{sys}} (\varpi/100 \, {\rm AU})^{-3/4}$.
It also shows the disk radii $R_{d,\lambda_{sys}}$ given by eq. (65) of S07, using the corresponding 
total disk mass in Table  \ref{table:1}. 
The models with $\lambda_{sys}=12$ have larger surface density (smaller disk radii) than the models with $\lambda_{sys}=4$
because the viscosity decreases and the mass accretion rate, proportional to $ \Sigma_{\varpi} \nu$,
is constant. 
The change in surface density has important implications on the disk structure as shown in
Figure \ref{fig:act-pass12} in comparison with the models in Figure \ref{fig:act-pass}. 
This figure shows the active and passive regions for the LMP, T Tauri, and FU Ori disk.
Because  the surface density increases by factors of $4 - 5$ in the LMP and the FU Ori disks, 
the active regions (yellow) are larger than the previous models, since the
stellar irradiation cannot penetrate much into the midplane. Instead, the density of the T Tauri disk increases only
by a factor of 1.5 while the aspect ratio $A_{12}$ increases by a factor of 8. Then,
the active region of the T Tauri disk decreases because the disk is more flared and intercepts more stellar flux.
Extensive modeling and spectra of magnetized disks around different YSOs, including 
an exploration of the parameter space  will be presented in a forthcoming paper (Tapia et al. in preparation).

 To calculate the vertical structure of the S07 radial models of magnetized disks we have assumed 
for simplicity a uniform viscous and resistive dissipation rate (eq. \ref{eq:dfluxvr}), independent of 
the height $z$.
One could assume instead  a viscosity $\nu(z)$ proportional to the local sound speed $a$, since
 in the S07 models the viscosity $\nu$
can also be written as a Shakura-Sunyaev viscosity $\nu = \alpha a^2/\Omega_K$ (eq. 49 of S07).
We also assumed that the viscous and resistive heating occur throughout the
disk, even though for the T Tauri disk the viscosity coefficient $D \sim 10^{-2.5}$  takes into account a
reduced efficiency in the viscous transport (S07).  Nevertheless, the models presented here can be
 modified to consider layered accretion, with viscous heating occurring only in the surface layers. 
 To do this in a self consistent way, one needs to 
 calculate the cosmic ray, X ray, thermal, and radioactive ionization 
to obtain the extent of the so-called ``dead zones'' in these magnetized  disk 
models (e.g., Umebayashi \& Nakano 2009; Cleeves, Adams \& Bergin 2013).
We leave this problem for future study.

{A relevant question is if the  S07 models of magnetized accreting disks threaded 
by a poloidal magnetic field are unstable to the MRI. 
Since the seminal papers on the MRI of Balbus \& Hawley (1991; hearafter BH91) and Hawley \& Balbus (1991),
several authors have studied the stability of different magnetized disks  models.
For example, Pessah \& Psaltis (2005) studied the 
stability of polytropic magnetized disks with superthermal toroidal fields, 
including  magnetic tension forces. The critical 
wavenumber for instability is modified with respect to the value of BH91.
They recovered the dispersion relation 
obtained by several authors in different limits: no field curvature (Blaes \& Balbus
1994); no compressibility (Dubrulle \& Knobloch 1993); and cold limit with no
field curvature (Kim \& Ostriker 2000), and made a thorough discussion of the
origin of the instabilities and approximations in the different regimes. 
Also, Ogilvie (1998) studied the stability of polytropic rotating disks threaded by
a poloidal magnetic field but without mass accretion. 
In particular, the S07 models satisfy  the general criterion that 
the ratio of the gas pressure to the  magnetic pressure $\beta > 1$.  This criterion obtained
by BH91, comes from the condition that the shortest wavelength unstable mode fits the vertical disk size.
For the local model considered by BH91, the normalized critical
wavenumber parameter for the MRI is 
$q_{\rm crit} ={k_{z,crit} v_{A,z} / \Omega} = \left\vert- 2 ({d \ln \Omega/ d \ln \varpi}) \right\vert^{1/2}$, which has the value 
$q_{\rm crit} = 3^{1/2}$ for Keplerian rotation.
Assuming a thermal disk scale height $H_{\rm thermal} = \sqrt{2} a / \Omega_K$, the critical wavelength 
divided by the disk size is 
${\lambda_{crit} /( 2 H_{\rm thermal})}= {\pi /( q_{crit} \beta^{1/2}}).$
Thus, for $\beta > 3^{1/2}$ the shortest wavelength unstable 
mode, $\lambda_{\rm crit}$,  fits the vertical disk size. 
Nevertheless, the S07 disk models are compressed by the poloidal magnetic field. From
their eq. (46)  the scale height of these magnetized disks can be written as $H = H_{\rm thermal} / c_B $ 
with $c_B > 1$. For computational purposes we choose to express the magnetic compression coefficient as
$c_B = \left[ 1+ I_l (1-f^2)/A\right]^{1/2}$, 
where $I_l$ gives the magnetic field inclination angle ( see Table 1 of S07),
and the aspect ratio $A = H/\varpi$, is obtained from the vertical structure models (see Table  \ref{table:2}). 
Thus, the ratio of the critical wavelength to the disk size is modified as
${\lambda_{crit} / (2 H)}= \pi c_B /( q_{crit} \beta^{1/2})$.
The value of the system mass-to-flux ratio $\lambda_{sys}$ determines the level of the 
disk magnetization and the value of the coefficient $c_B$.
The strongly magnetized models with mass-to-flux ratio
$\lambda_{sys}=4$ at 1 AU have $c_B= 2.0, 15.5, 6.9$ for the LMP, T Tauri, and LMP disks,
respectively. The less magnetized $\lambda_{sys} =12$ models at 1 AU
have $c_B = 1.1, 2.1, 1.4$ for the LMP, T Tauri, and LMP disks, respectively. 
The magnetic correction  decreases slowly with radius  $I_l (1-f^2)/A \propto (\varpi/ AU)^{-1/4}$,
and $c_B \rightarrow 1$. If $c_B >>1$, one would expect the disk to be stable to the MRI.
Instead, the low values of $c_B$ for the models discussed in this work suggest that these disks can sustain the MRI, 
except  maybe in the case of the strongly magnetized T Tauri disk with $\lambda=4$.
Table \ref{table:3} show the values of the 
Elsasser number $A_m = v_A^2/(\eta \Omega)$ and  plasma $\beta = 2 a^2/v_A^2$, where the Alfven speed is 
$v_A = B/ \sqrt{4 \pi \rho}$, for the S07 models at the disk midplane at 1 AU.
The Elsasser number decreases slowly with radius as $A_m = A_{m,1}( \varpi/ 1 AU)^{-1/4}$. The gas to magnetic pressure 
increases slowly with radius as $\beta =\beta_1(\varpi/1 AU)^{1/4}$ for weakly magnetized disks, and is constant for strongly 
magnetized disks. Thus, the S07 models have both $A_m >1$ and $\beta > 1$,  indicating that the disks should
unstable to the MRI. Nevertheless, one requires a detailed stability analysis of the S07 models, which is beyond of the scope
of this paper and we leave as future work.  }

{As mentioned in the Introduction, several authors have made simulations of magnetized disks
including Ohmic resistivity, ambipolar diffusion and the Hall effect, and find that the MRI is 
suppressed and that a disk wind transports away the angular momentum transport driving the
disk accretion is (e.g., Bai 2015;  Gressel et al. 2015, without the Hall effect).   These simulations consider weak magnetic
fields in the sense that $\beta \sim 10^{4-5}$. In contrast, the S07 radial models have low values of 
$\beta < 100$ throughout the disk, as shown in Table  \ref{table:3}. It would be important to study
magnetized disk simulations in the parameter regime of the S07 models, that would also take 
into account the back reaction of the magnetic field on the flow. Such simulations would help 
determine if the MRI operates in the S07 models or if a basic assumption of these models, 
hydrostatic equilibrium, is flawed.
}

{ Finally, the vertical structure of the irradiated magnetized disk models studied in this work can be compared 
to other disk models of the vertical structure of nonmagnetic accretion disks commonly used in the literature.
 For example, Chiang \& Goldreich (1997) modeled passive disks with 2 zones: an upper layer where the dust is heated by
by the stellar radiation and a midplane region heated by radiation reprocessed by the dust in the upper layer.
The interior is isothermal and in hydrostatic equilibrium. 
These models are semi-analytic and easy to implement but they do not solve for the vertical temperature gradients that one 
can see in Figures (1-3). Also, they are passive disk models, so they do not  include viscous heating.
On the other hand, the models of D'Alessio et al. (1998)  solve for the temperature gradients and include both stellar
irradiation and viscous heating. The models  presented in this work also solve for the temperature gradients and include irradiation, 
viscous, and resistive heating. Furthermore, the magnetic field compression can be important in the strongly magnetized
disks, reducing their aspect ratio with respect to the nonmagnetic disk models. }

 \section{Conclusions}
 
This work presents the first models of the vertical structure of irradiated magnetized accretion disks threaded by a 
poloidal magnetic field dragged in during the process of disk formation.
 These disks are subject to viscous and resistive heating and to irradiation by the central star. 
 We calculate the vertical structure of disks around LMP, FU Ori, and T Tauri stars.
 We use the radial models of S07, although our formalism
 can be applied to other models of the disk radial structure.
 
We considered strongly magnetized disks with a mass-to-flux ratio $\lambda_{\rm sys} = 4$ and the parameters of the standard
models of S07 in Table  \ref{table:1}. We find that the T Tauri disks 
are compressed by the magnetic pressure and are very thin and cold compared with observations. 
The LMP disk midplane thermal structure is dominated by viscous heating while 
the FU Ori disk midplane thermal structure is dominated by resistive heating. The T Tauri disk midplane has similar contributions
for the viscous and resistive heating
 
 Changing the mass-to-flux ratio $\lambda_{\rm sys}$ changes the disk structure. In particular, we 
 considered a larger value $\lambda_{\rm sys} = 12$ (less magnetization) which increases the disk density 
 and decreases the magnetic compression. In these models, the T Tauri disk has a larger aspect ratio, 
 consistent with observations. In all these disk viscous heating dominates the midplane
 thermal structure since resistive heating is less than 7\% of the viscous heating. Also, the size of the active region 
 in the weakly magnetized disks changes depending on which effect dominates:  the disk becomes denser and the
 irradiation cannot penetrate increasing the active region in the LMP and FU Ori disks, 
 or the disk is more flared and intercepts more irradiation 
 decreasing the active region in the T Tauri disk.
  
Surface irradiation by the central source produces hot atmospheres in the disks.
These atmospheres (above and below the midplane) are spatially extended but contain little mass,
less than 10\%  of the total mass surface density. 
The disks show large midplane areas with vertical temperature inversions from the midplane up to the base of 
the hot amosphere. 

Finally, the radial and vertical structure of magnetized accretion disks around young stars
 and the importance of the resistive heating depend on 
 the system mass-to-flux ratio $\lambda_{sys}$. One expects that in the near future 
 ALMA will be able to measure magnetic fields and their morphology in protoplanetary disks directly 
through Zeeman splitting of the CN molecule and polarization of dust emission. 
Such measurements, together with the disk mass will determine observationally the relevant values of $\lambda_{sys}$,
and thus, the importance of magnetic fields in the disk evolution, structure and emission.

 \acknowledgments

We thank an anonymous referee for very useful comments and suggestions that improved the presentation of this
paper.
SL, CT and YB acknowledge support by  CONACyT 153522/238631 and DGAPA-UNAM IN100412/IN105815.

\appendix
\section{Appendix: Transport equations}

For simplicity, we assume below that the energy flux is only 
radiative, i. e.,  convective and conductive fluxes will be ignored. Nevertheless, the Schwarzschild stability criterion 
for convection is checked at each step of point and, if necessary,  the gradient can be modified following Mihalas (1978). In the
few zones where convection occurs, we find that the temperature gradient is very close to the radiative gradient.

\subsection{Non-irradiated Viscous and Resistive  Disk}

Consider a viscous and resistive disk without irradiation from the central star.
Assuming a uniform dissipation rate ( eq. \ref{eq:internal_flux}), the vertical energy flux is
\be
F_{ vr}(\Sigma) = F_{vr,\varpi}\frac{\Sigma}{\Sigma_\varpi},
\label{eq:fluxvr}
\ee
where $\Sigma$ is the midplane surface density defined by
eq. (\ref{Sigmamid}).
The vertical flux equation is 
\be
\frac{d F_{vr}}{d \Sigma} = \frac{ F_{vr,\varpi}}{\Sigma_\varpi} = \frac{3}{2} \frac{\nu}{\Sigma_\varpi} \left( \frac{3}{2} f^2 \Sigma_\varpi \Omega_K^2 + 
\frac{1}{\varpi} \frac{ B_\varpi^+ B_z}{2 \pi}  \right),
\label{eq:dfluxvr}
\ee
where the total viscous resistive flux $F_{ vr,\varpi}$ 
emerges from both faces of the disk surface (half on each surface).

The frequency integrated first moment of the 
transport equation gives the algebraic equation 
\be
 \frac{\sigma T_{vr}^4}{\pi} = J_{vr} + \frac{1} {4 \pi \kappa_P}\frac{d F_{vr }}{d \Sigma} ,
\label{Tvr}
\ee
where
$\kappa_P $ is the Planck mean opacity, $\sigma$ is the Stephan-Boltzmann constant, and $T_{vr}$ and
$J_{vr}$ (${\rm erg\, cm^{-2} s^{-1} Hz^{-1} sr^{-1}}$) are the temperature and mean intensity of the viscous resistive disk, respectively. The 
frequency integrated, second moment of the 
transport equation plus the Eddington approximation give the equation for the mean intensity $J_{vr}$, 
\be
\frac{d J_{vr}}{d \Sigma} = -\frac{3 \chi_R }{4 \pi} F_{vr}, 
\label{dJvr}
\ee
where $\chi_R$ is the Rosseland mean opacity.
From the two stream approximation, the boundary condition is know at the upper boundary
\be
J_{vr}({\Sigma_\varpi/ 2}) = {\sqrt{3} \over 4 \pi} {F_{vr,\varpi} \over 2},
\label{eq:BCvr}
\ee
where $\Sigma_\varpi/2 $ is half the total surface density at the disk surface $z_\infty$, and 
$F_{vr,\varpi} / 2$ is the flux that exits from that surface. 

Eqs. (\ref{Tvr}) and (\ref{dJvr}) can be solved for the  viscous resistive disk temperature $T_{vr}$ and mean intensity $J_{vr}$.

\subsection{Irradiated Passive Disk}

Consider a passive disk irradiated by the central star.
We follow D'Alessio et al. (1998)
who assumed that the radiation field is separated into two components: the ``stellar" 
and the ``disk", where the external irradiation is characterized by wavelengths 
different from those of the local disk radiation field, 
as first proposed by Strittmatter (1974) in the case 
of  irradiation by X-rays on the atmosphere of a close binary.
Thus, the transfer equations are solved for the disk and the stellar scattered radiation using
different mean opacities: for the stellar irradiation one uses a ``true'' absorption
coefficient $\kappa_P^s$ and a Planck average
 mean extinction coefficient $\chi_P^s$ , where the Planck function is evaluated at the temperature of
 the central source, $B(T_s)$;
  and for the disk radiation, one uses a ``true'' absorption coefficient $\kappa_P$, 
and a Rosseland mean opacity $\chi_R$, and where the Planck function is evaluated at the
disk temperature, $B(T)$. Here we will consider the radiation field of the central source at an
effective irradiation temperature $T_s = T_{eff}$, that takes into account the total bolometric luminosity, 
stellar plus accretion luminosity.

The disk intercepts at the upper boundary a stellar flux, $F_{irr}$, at an angle, $\mu_0$, given
by the shape of the disk surface as discussed in Appendix B. The stellar radiation optical depth from the height 
$z$ to the disk surface $z_\infty$ is 
\be
\tau_s = \int_{z}^{z_\infty} \rho \chi_P^s dz = \int_{\Sigma}^{\Sigma_\varpi/2} \chi_P^s d \Sigma.
\label{taus}
\ee

The scattered stellar radiation is characterized by a
mean intensity and scattered flux given by 
\be
J_{scatt} = \sigma_s \frac{F_{irr}}{4 \pi} \left \{ 
\frac {3( 1+ \sqrt{3}  \mu_0)}{ [\sqrt{3} + \beta_s] (1-\beta_s^2 \mu_0^2)} 
e^{-\beta_s \tau_s} - \frac{3 \mu_0}{(1-\beta_s^2 \mu_0^2)}
e^{-\tau_s/\mu_0} \right \},
\ee  
\be
F_{scatt} = \sigma_s F_{irr} \left \{ -
\frac {\beta_s (1 + \sqrt{3} \mu_0)}{[ \sqrt{3} + \beta_s] 
(1-\beta_s^2 \mu_0^2)} e^{-\beta_s \tau_s} + \frac{1}{(1-\beta_s^2 \mu_0^2)}
e^{-\tau_s/\mu_0} \right \},
\label{Fs}
\ee 
where the albedo is $\sigma_s = (\chi_P^s - \kappa_P^s)/\chi_P^s$, the fractional absorption is
\be
a_s=\kappa_P^s/\chi_P^s,
\label{eq:frac_abs}
\ee
and $\beta_s=\sqrt{3 a_s}$.

For a passive irradiated disk, the zero flux condition in the vertical direction gives the reprocessed flux,
\begin{eqnarray}
F_{rp} &=&  F_{irr} e^{-\tau_s/\mu_0} - F_{scatt} ,\nonumber \\
        &=&F_{irr} a_s [ (1+C_1) e^{-\tau_s/\mu_0} + C_2 e^{-\beta_s \tau_s} ],
\label{Frp}
\end{eqnarray}
where $C_1$ and $C_2$ are given by
\be
C_1 = -{\frac{ 3 \sigma_s \mu_0^2}{1 -\beta_s^2\mu_0^2}}
\quad {\rm and} \quad
C_2 = 
{\frac{ 3 \sigma_s (1+\sqrt{3}  \mu_0)}{\beta_s (\sqrt{3}+ \beta_s)(1-\beta_s^2\mu_0^2)}}.
\label{C12}
\ee
The frequency integrated first moment of the disk radiation transfer equation gives
\be
 \frac{\sigma T_{rp}^4}{\pi}= J_{rp} + \frac{1}{4 \pi \kappa_P} \frac{d F_{rp}}{d\Sigma}
= J_{rp} - \frac{\chi_P^s}{4 \pi \kappa_P} \frac{d F_{rp}}{d\tau_s}, 
\label{Trp} 
\ee
where $J_{rp}$ is the mean intensity (${\rm erg\, cm^{-2} s^{-1} Hz^{-1} sr^{-1}} $)  of the disk reprocessed radiation.
To obtain the last equality on the RHS,
 which is useful for computational purposes (see eq. \ref{Frp}), we have substituted
$d \tau_s = - \chi_P^s d \Sigma$.

Also, from frequency integrated, second moment of the disk radiation transfer equation 
and the Eddington approximation one gets an equation for the mean intensity, $J_{rp}$,
\be
\frac{d J_{rp}}{d\Sigma} = - \frac{3}{4 \pi} \chi_R F_{rp}.
\label{dJrp}
\ee
The  boundary condition is obtained from the  two-stream approximation, 
\be
J_{rp}(\Sigma_\varpi/2)=
{\sqrt{3} \over 4 \pi} F_{rp}(\tau_s=0) = {\sqrt{3} F_{irr} a_s \over  4 \pi} \left( 1 + C_1 + C_2 \right).
\label{eq:BCrp}
\ee 

Finally, eqs. (\ref{Trp}) and (\ref{dJrp}) can be solved for the  disk temperature $T_{rp}$ and mean intensity $J_{rp}$
of the reprocessed stellar radiation in the disk.

\section{Disk Irradiation}
\label{sect:irrad}

\begin{figure}
\centering
  \includegraphics[angle=0,width=0.95\textwidth]{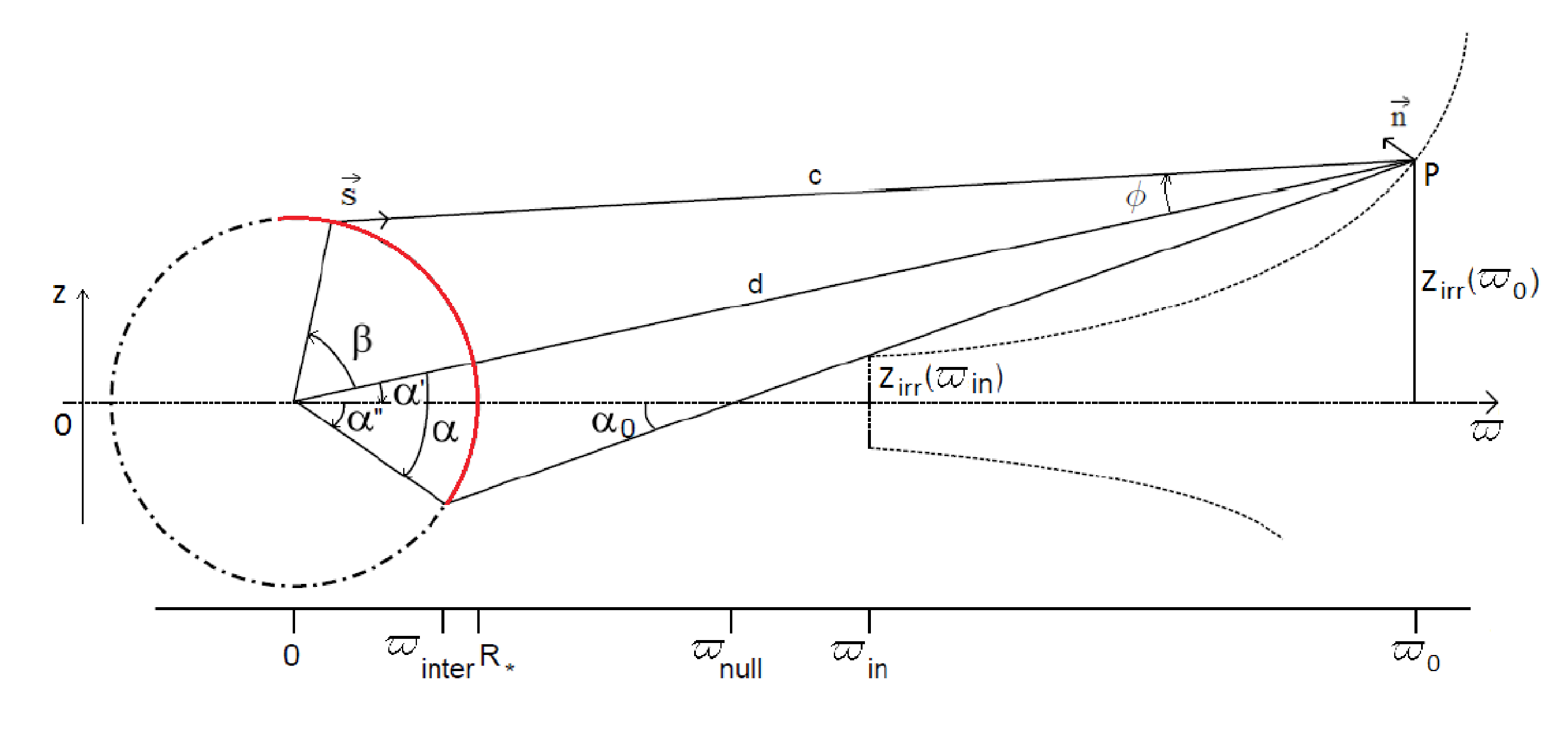}
  \caption{Representation of a flared disk. The disk surface and midplane are shown as dashed lines. 
  The stellar surface is plotted as a dot-dashed line and the red solid line indicates the surface visible 
  from point P.}
  \label{fig:disk-profile}
\end{figure}

The calculation of the stellar irradiation on the disk surface follows the treatment of 
Kenyon \& Hartmann, 1987 (hereafter HK87).  A cartoon of the star plus disk system is presented on the 
Figure \ref{fig:disk-profile} and 
uses the same notation as in HK87 when possible. We assume that the 
 disk is truncated at an internal radius $\varpi_{in}$, at which the dust sublimates ($ T \sim 1,400$ K).
 The irradiation surface  $z_{irr}(\varpi)$  is obtained numerically, integrating the optical depth $\tau_*$ starting from
rays originating at the star center up to the point $\tau_* = 1$, i.e., where the stellar radiation is absorbed.

Once the irradiation surface is determined, one has to obtain the stellar flux at each point on this surface 
$P=[\varpi_0,z_{irr}(\varpi_0)]$ . 
Let us call the line connecting $P$ with the centre of the star the ``symmetry line'', indicated in 
Figure \ref{fig:disk-profile}. This line has a length $d$.
The radiation received on $P$ is calculated integrating over concentric annuli on the
stellar surface with angular  radius $\phi$ centered around the symmetry line, and
angular length $\theta$ measured on the plane normal to the symmetry line, 
\begin{equation}
    F_{irr}(P) = 2 \int_{0}^{\phi_{max}} \int_{0}^{\theta_{max}} I \sin\phi   \, \vec{s} \cdot (-\hat n)
    \,d\theta \,d\phi ,
\label{eq:flux-1}    
\end{equation}
where $I= L_c/ 4 \pi^2 R_*^2$  is the specific intensity from the central source  with  units 
\{${\rm erg \, cm^{-2} \, s^{-1}  \, str^{-1}}$\}, and 
$L_c$ is the central source luminosity (accretion plus stellar) and $R_*$ is the stellar radius. 
The unit vector $\hat n $, normal to the irradiation surface, is 
\begin{equation}\label{eq:normal}
  \hat n = \left .  \frac{-z_{irr}(\varpi_0)^\prime \hat{\varpi}+\hat z}{\left[1+ \left[ z_{irr}(\varpi_0)^\prime\right]^{2}\right]^{1/2}}  \right|_P,
\end{equation}
where $z_{irr}(\varpi)^\prime = d z_{irr}(\varpi)/ d\varpi$, and $\hat \varpi$ and $\hat z$ are the unit vectors in the
radial and vertical direction, respectively. 
The vector $\vec{s}$ from each annulus on the star to the point $P$ is given by
\begin{eqnarray}
   \vec s & = & \left [ \frac{\varpi_0}{d} \cos\phi + \frac{z_{irr}(\varpi_0)}{d} \sin\phi  \cos\theta \right ] \hat \varpi
                +\sin\phi \, \sin\theta (\hat z \times \hat \varpi) \\ \nonumber
            & & + \left [ \frac{z_{irr}(\varpi_0)}{d} \cos\phi - \frac{\varpi_0}{d} \sin\phi  \cos\theta \right ] \hat z .
   \label{eq:surface}
\end{eqnarray}
Then, the irradiation flux  given by eq. (\ref{eq:flux-1}) can be written as 
\begin{equation}
    F_{irr}(P) = 
    2 \int_{0}^{\phi_{max}} \sin\phi \,d\phi \int_{0}^{\theta_{max}}  I
    \frac{A_1 \sin\phi \cos\theta + A_2 \cos \phi}{A_3}\,d\theta,
    \label{eq:flux-2}
\end{equation}
where 
\begin{eqnarray}
        A_1& = &\varpi_0 + z_{irr}(\varpi_0) z_{irr}(\varpi_0)^\prime, \\ \nonumber
        A_2& = & \varpi_0 z_{irr}(\varpi_0)^\prime - z_{irr}(\varpi_0), \\ \nonumber
        A_3&= & d \left ( 1 + \left[z_{irr}(\varpi_0)^\prime\right]^2 \right )^{1/2} .
      \label{eq:c1-c2}
\end{eqnarray}
The maximum value
\be
\phi_{max} = \sin^{-1}\left({R_* \over d}\right),
\ee
gives the stellar angular size seen from $P$. 
The upper limit $\theta_{max}$  determines which part of each annulus is visible.
To calculate this limit, one needs to determine if the inner disk hides
part of the star. Consider 
the lowest l.o.s. from $P$ to the star that has the equation
\begin{equation}\label{eq:sightline}
    z_{lowest}= (\varpi-\varpi_{null}) \tan(\alpha_0),
\end{equation}
where the angle with respect to the midplane is
\begin{equation}\label{alpha0}
    \alpha_0 = \arctan\left({ {z_{irr}(\varpi_0)- z_{irr}(\varpi_{in})} \over \varpi_0-\varpi_{in} }\right).
\end{equation}
and $\varpi_{null}$ is the radius at which this line of sight intersects the midplane,
\begin{equation}\label{R_null}
    \varpi_{null} = \varpi_{in}- {z_{irr}(\varpi_{in}) \over \tan(\alpha_0)}.
\end{equation}
On the other hand, the equation of the surface of the star is
\begin{equation}\label{ystar}
    z_* = \pm \sqrt{R_*^2 - \varpi^2}, 
\end{equation}
where the sign is positive (negative) above (below) the midplane. The intersection radius $\varpi_{inter}$ 
is given by  $z_{lowest} = z_*$ which  leads to a second degree equation for $\varpi_{inter}$,
\begin{equation}
    \left[\tan(\alpha_0)^2+1\right]\varpi_{inter}^2 - 2\varpi_{null} \tan(\alpha_0)^2 \varpi_{inter}+
    \varpi_{null}^2 \tan(\alpha_0)^2-R_*^2 = 0.
\label{eq:second}
\end{equation}
When this equation has no real solutions it 
mean that the lowest l.o.s.  does not intersect the stellar surface. If $\varpi_{null} > R_*$, 
which is the most common case, all the stellar surface is visible from $P$, and $\theta_{\rm max} = \pi$. 
Otherwise, the star is hidden by the inner disk and $\theta_{\rm max} = 0$.
When eq. (\ref{eq:second}) has two real solutions, they correspond to the line 
$z_{lowest}$ intersecting the stellar surface twice.
One needs to consider only the larger value of $\varpi_{inter}$, closest to $P$. 

Let $\alpha^\prime$ be the angle between 
 the symmetry line and the disk midplane, given by
 \begin{equation}\label{eq:alphap}
  \alpha^{\prime} = \arctan{  \left({z_{irr}(\varpi_0)\over \varpi_0}\right )  }.
\end{equation}
Also,  let $\alpha^{\prime\prime}$ be the angle measured  between the midplane and the lowest 
l.o.s.
\be
\alpha^{\prime\prime}= \left\{
\begin{array}{ll}
\arccos({\varpi_{inter} \over R_*}), & \varpi_{null} > R_* ,\\
-\arccos({\varpi_{inter} \over R_*}), & \varpi_{null} < R_* .
\end{array}
\right .
\ee
In the first case,  the intersection occurs in the lower stellar hemisphere;
in the second case, it occurs in the upper hemisphere.
The total angle $\alpha$ is given by
\begin{eqnarray}
  \alpha &=& \alpha^{\prime} + \alpha^{\prime\prime} \\  \nonumber
         &=& \arctan{\left( {z_{irr}(\varpi)\over \varpi} \right)}\pm \arccos{\left( {\varpi_{inter}\over R_*}\right)}.
\end{eqnarray}
Now, given the annulus observed from $P$ with angular size $\phi$, let $\beta$ be the angular radius 
measured from the star center
\be
\beta = \arcsin{ c \sin \phi \over R_*}, 
\ee
where the length of the line connecting $P$ and the annulus is $c= d \cos\phi - \sqrt{R_*^2 - d^2 \sin^2\phi}$.

Finally, given the angles $\alpha$ and $\beta$, the upper limit $\theta_{max}$  is  
\be
\theta_{max} = \left\{
\begin{array}{ll}
\pi ; &   \alpha < \beta ,  \\
 \pi-\arccos{\left[\sin(\alpha)\over \sin(\beta) \right]} ; &   \alpha \, \in \, [-\beta,\beta], \\
 0 ; &   \alpha  < -\beta. 
\end{array}
\right .
\ee
In the first case, $\theta_{max} = \pi$ means that all the annulus of angular radius 
$\phi$ is seen from $P$. In the second case, a part of the star is hidden by the disk. 
In the last case, $\theta_{\rm max} = 0$ means all the annulus is hidden.

 \begin{deluxetable}{llllllll}
\tablecolumns{8}
\tablewidth{0pc}
\tablecaption{Model Parameters}
\tablehead{
 \colhead{YSO} & \colhead{$\dot M_{d} $} & \colhead{$D$} & \colhead{$M_d$}  
 & \colhead{$f$ } & \colhead{$R_*$} & \colhead{$L_c $}&  \colhead{$T_{eff}$} \\
 & $(M_\odot \yr^{-1})$ &  &$(M_\odot)$ &   &   $(R_\odot)$ &  $(L_\odot)$ & (K) 
  }
\startdata
LMP & $2\times 10^{-6}$ & 1  & 0.2 &0.957 &  3 &  7.1 & 5490\\ 
T Tauri  & $1\times 10^{-8}$ & $10^{-2.5}$  &  0.03 & 0.658 &  2 & 0.93 & 4040 \\
FU Ori & $2\times 10^{-4}$ & 1 &  0.02  & 0.386 &  7 & 230 & 8570 
\enddata 
\label{table:1}
\end{deluxetable}

 \begin{deluxetable}{lllllllll}
\tablecolumns{7}
\tablewidth{0pc}
\tablecaption{Models with different mass-to-flux ratio $\lambda_{sys}$ }
\tablehead{
 \colhead{YSO} & \colhead{$f_4$ } & \colhead{$A_{4}$} & $\Sigma_{4}$ & $R_{d,4}$ &  \colhead{$f_{12}$}&  \colhead{$A_{12}$} & $\Sigma_{12} $ & $R_{d,12}$ \\
 & & & $ {\rm g/cm^2}$ & AU   & & & ${\rm g/cm^2}$ & AU }
\startdata
LMP &  0.957 & 0.156 & 5.33 & 455 &  0.995 & 0.309 &  25.2 & 131 \\ 
T Tauri  & 0.658 & 0.013 & 10.4 & 58.2 &  0.968 & 0.109 & 16.5 & 40.4 \\
FU Ori & 0.386 & 0.102 & 33.0 & 16.8 & 0.952  & 0.525 & 142 & 5.22
\enddata 
\label{table:2}
\tablecomments{The disk aspect ratio is $A(\varpi) = A_{\lambda_{sys}} (\varpi/100 \, {\rm AU})^{1/4}$. 
The surface density is  $\Sigma(\varpi)  = \Sigma_{\lambda_{sys}} (\varpi/100 \, {\rm AU})^{-3/4}$. }
\end{deluxetable}

 \begin{deluxetable}{llll}
\tablecolumns{8}
\tablewidth{0pc}
\tablecaption{Elsasser number $A_m$ and plasma $\beta$ at 1 AU}
\tablehead{
 \colhead{YSO} & \colhead{$\lambda_{sys}$} & \colhead{$A_{m,1}$} & \colhead{$\beta_1$} \\ 
  }
\startdata
LMP & 4 & 22.5 & 3.50 \\
        & 12 & 11.8 & 20.5\\
T Tauri  & 4 & $5.88\times 10^{4}$ & 2.50\\
            & 12 & $1.03 \times 10^{4}$ & 3.44\\
FU Ori & 4 & 13.9  & 2.56  \\
            & 12  & 6.66  & 5.55
\enddata 
\label{table:3}
\end{deluxetable}

 \end{document}